\documentclass[final]{agujournal}
\draftfalse

\journalname{Space Weather}

\usepackage{amssymb}
\usepackage{amsthm,amsmath}
\DeclareMathOperator\erf{erf}

\begin{document}

%% ------------------------------------------------------------------------ %%
%  Title
%
% (A title should be specific, informative, and brief. Use
% abbreviations only if they are defined in the abstract. Titles that
% start with general keywords then specific terms are optimized in
% searches)
%
%% ------------------------------------------------------------------------ %%

% Example: \title{This is a test title}

\title{On the generation of probabilistic forecasts from deterministic models}

%% ------------------------------------------------------------------------ %%
%
%  AUTHORS AND AFFILIATIONS
%
%% ------------------------------------------------------------------------ %%

% Authors are individuals who have significantly contributed to the
% research and preparation of the article. Group authors are allowed, if
% each author in the group is separately identified in an appendix.)

% List authors by first name or initial followed by last name and
% separated by commas. Use \affil{} to number affiliations, and
% \thanks{} for author notes.
% Additional author notes should be indicated with \thanks{} (for
% example, for current addresses).

% Example: \authors{A. B. Author\affil{1}\thanks{Current address, Antartica}, B. C. Author\affil{2,3}, and D. E.
% Author\affil{3,4}\thanks{Also funded by Monsanto.}}

\authors{E. Camporeale\affil{1}, X. Chu\affil{2}, O.V. Agapitov\affil{3}, J. Bortnik\affil{4}}

% \affiliation{1}{First Affiliation}
% \affiliation{2}{Second Affiliation}
% \affiliation{3}{Third Affiliation}
% \affiliation{4}{Fourth Affiliation}

\affiliation{1}{Center for Mathematics and Computer Science (CWI), Amsterdam, The Netherlands}
\affiliation{2}{Laboratory for Atmospheric and Space Physics, University of Colorado, Boulder, CO, USA}
\affiliation{3}{Space Sciences Laboratory, University of California Berkeley, Berkeley, CA, USA}
\affiliation{4}{Department of Atmospheric and Oceanic Sciences, University of California, Los Angeles, CA, USA}

%(repeat as many times as is necessary)

%% Corresponding Author:
% Corresponding author mailing address and e-mail address:

% (include name and email addresses of the corresponding author.  More
% than one corresponding author is allowed in this LaTeX file and for
% publication; but only one corresponding author is allowed in our
% editorial system.)

% Example: \correspondingauthor{First and Last Name}{email@address.edu}

\correspondingauthor{Enrico Camporeale}{e.camporeale@cwi.nl}

%% Keypoints, final entry on title page.

% Example:
% \begin{keypoints}
% \item	List up to three key points (at least one is required)
% \item	Key Points summarize the main points and conclusions of the article
% \item	Each must be 100 characters or less with no special characters or punctuation
% \end{keypoints}

%  List up to three key points (at least one is required)
%  Key Points summarize the main points and conclusions of the article
%  Each must be 100 characters or less with no special characters or punctuation

\begin{keypoints}
\item We introduce a new method to estimate the uncertainties associated with single-point outputs generated by a deterministic model
\item The method ensures a trade-off between accuracy and reliability of the generated probabilistic forecasts
\item The method is computationally inexpensive, avoiding the costs associated with ensemble simulations, and is model independent. The only inputs needed are the observed errors between predictions and observations of the deterministic model.
\end{keypoints}

%% ------------------------------------------------------------------------ %%
%
%  ABSTRACT
%
% A good abstract will begin with a short description of the problem
% being addressed, briefly describe the new data or analyses, then
% briefly states the main conclusion(s) and how they are supported and
% uncertainties.
%% ------------------------------------------------------------------------ %%

%% \begin{abstract} starts the second page

\begin{abstract}
Most of the methods that produce space weather forecasts are based on deterministic models. In order to generate a probabilistic forecast, a model needs to be run several times sampling the input parameter space, in order to generate an ensemble from which the distribution of outputs can be inferred. However, ensemble simulations are costly and often preclude the possibility of real-time forecasting.\\
We introduce a simple and robust method to generate uncertainties from deterministic models, that does not require ensemble simulations. The method is based on the simple consideration that a probabilistic forecast needs to be both accurate and well-calibrated (reliable). We argue that these two requirements are equally important, and we introduce the Accuracy-Reliability cost function that quantitatively measures the trade-off between accuracy and reliability. We then define the optimal uncertainties as the standard deviation of the Gaussian distribution that minimizes the cost function.\\
We demonstrate that this simple strategy, implemented here by means of a regularized deep neural network, produces accurate and well-calibrated forecasts, showing examples both on synthetic and real-world space weather data.
\end{abstract}

%% ------------------------------------------------------------------------ %%
%
%  TEXT
%
%% ------------------------------------------------------------------------ %%

%%% Suggested section heads:
\section{Introduction}

The US National Space Weather Action Plan released in October 2015 has fueled interest in so-called Operations-to-Research (O2R) activities, which are now explicitly funded by NASA and NOAA programs.
An important element of O2R is the enhancement of existing operational models and products with fundamental research. A major weakness of most of the state-of-the-art forecasting models used by national Space Weather agencies is that they are essentially deterministic. For any given set of input parameters, they output a single-point estimate, without providing information on the uncertainty associated with such an estimate.
On the other hand, the Space Weather community is gradually recognizing the importance of probabilistic forecasts, which have been the standard in meteorological weather forecast for many years.
Indeed, several probabilistic forecasting models have been proposed in the last few years, concerning solar energetic particles \citep{kahler15, aminalragia18}, geomagnetic indexes \citep{mcPherron04, zhang14,riley16, chandorkar17,gruet18}, GPS scintillation \citep{prikryl2012}, solar flares \citep{gallagher02, barnes03, wheatland04, lee12, bloomfield12, papaioannou15}, solar wind speed \citep{Bussy‐Virat14, owens17, napoletano18}, and relativistic electron fluxes \citep{miyoshi08}, among others.\\
As pointed out in \citet{Murray17}, most operational space weather forecasting centers worldwide still rely on human forecasters to adjust the issued probability of a given event, based on experience. Yet, a recent verification of geomagnetic storm and X-ray flare forecasts issued by the Met Office Space Weather Operations Centre (MOSWOC), has reported that these forecasts struggle to provide a better prediction than a reference model, and tend to overforecast events \citep{Sharpe17}. Moreover, comparing eleven different methods to predict flares, \citet{barnes16} concluded that no participating method proved substantially better than climatological forecasts, for M-class flares and above (a climatological model is one where a long-term average of the quantity of interest is taken as forecast).\\ 
{There are two major approaches in producing a probabilistic model}. The first way is to apply a statistical method on historical records, trying to correlate some input parameters with the forecast output. Little or no physics assumptions enter in such models (other then maybe a judicious choice of input parameters based on physics). For instance, modern machine learning algorithms, often referred to as \emph{black-box} models, fall in this category \citep{ghahramani15, murphy12, camporeale18, camporeale18b}. 
A second way of producing a probabilistic forecast is based on the use of physics-based models, which range from (almost) first-principle simulations \citep[e.g.,][]{Luhmann17}, to semi-empirical models \citep[e.g.,][]{mostl17}. These \emph{white-box} models are typically deterministic, meaning that they return a single solution for any given set of inputs provided. How to assign a probabilistic interpretation to such single-point estimates, in a computationally cheap way, is a challenging open problem which forms the core of a research area called non-intrusive Uncertainty Quantification \citep{smith13}. Non-intrusive refers to the fact that one employs a deterministic model (and its existing software), and performs an ensemble of simulations, without changing the underlying equations. It is then straightforward to extract a probabilistic description from the results ensemble. However, this is usually very expensive, and brings the two following problems. First, if the number of inputs is large, one encounters the infamous \emph{curse of dimensionality}, namely the fact that the volume of an hypercube increases exponentially with the number of dimensions. Hence, sampling the input space with a tensorial grid (i.e. with a given number of points per dimension) quickly becomes unfeasible, because each grid point corresponds to a single run of a deterministic simulation. For this reason, sampling is often done in a Monte-Carlo fashion (or one of its modifications, such as Quasi-Monte-Carlo \citep{caflisch98}), which is very robust but also very slow in achieving convergence. Not surprisingly, an active area of research focuses on the design of adaptive sampling algorithms that yield convergence faster than Monte-Carlo \citep{xiu02, babuvska07, xiu10, camporeale17}. 
The second problem is that the distribution of outputs collected from the ensemble of simulations (the probabilistic forecast) is obtained by mapping, through the nonlinear simulation, the probability density that is assumed for the input parameters. Any misfit in the distribution of the inputs propagates to the distribution of outputs, producing misleading results. For this reason, an essential step of ensemble simulations is the calibration of the model \citep{kennedy01}, that is the derivation of the distribution of the input parameters that is most consistent with observations. {Calibration can itself be rather expensive, when it also relies on a large number of simulation runs}.\\

In this paper we introduce a new method to derive a probabilistic forecast based on a deterministic model, that avoids the computational costs associated with collecting an ensemble, and with properly calibrating a computer simulation.
We focus on the prediction of a continuous real variable. 
Hence, the method produces a probabilistic forecast in terms of a probability density function (pdf) for the quantity of interest. Moreover, we restrict ourselves to the case where such probability density is by construction Gaussian.\\

{\subsection{Accuracy and Reliability}}

This method is based on the simple consideration that a probabilistic forecast needs to be both accurate and reliable. This is in line with \citet{gneiting07}, that have proposed to evaluate the performance of a forecast based on the paradigm of maximizing the sharpness of the predictive distributions subject to calibration. {Sharpness refers to the concentration of the predictive
distributions and is a property of the forecasts only.} Note that in this paper we refer to calibration and reliability interchangeably, the former term typically being used in meteorological literature.  Following the seminal paper by \citet{murphy92}, accuracy is defined as the overall degree to which forecasts correspond to observations. It can be quantified introducing a proper scoring rule \citep{brocker07}, whose examples are the Brier score for binary events \citep{brier50}, the Rank Probability Score for multi-category events, and its generalization for forecast of continuous variables, the Continuous Rank Probability Score (CRPS) \citep{hersbach00, wilks11}, that we will use here. 
Reliability is the property of a probabilistic model that measures its statistical consistency with observations. In particular, for forecasts of discrete events, the reliability measures if an event occurs on average with frequency $p$, when it has been predicted to occur with probability $p$.  
{For example, consider a probabilistic, binary, meteorological model that predicts rain or no-rain. Take a large enough sample of predictions of '70\% chance of rain'. The model is said to be reliable/calibrated if approximately 70\% of these predictions turned out to be true (i.e., it rained), and if this holds for all forecasted probabilities. }
The same concept can be extended to forecasts of a continuous scalar quantity by examining the so-called rank histogram \citep{anderson96, hamill97,hamill01} or the reliability diagram \citep{pinson10}. 
{A reliability diagram represents, for any value of probability predicted for a given output, what is the actual observed frequency for that output (i.e., How many times did it rain, when 70\% chance of rain was predicted? More (under-confident), less (over-confident), or exactly 70\% of the time?).
In the case of continuous variables, the reliability diagram is obtained with the following straightforward procedure. One collects a (large) number of pairs observations-predictions (the former being a real number, the latter a probability density). For each observation, one computes what was the probability that was assigned to the the outcome being less or equal than the observed outcome. In the case of Gaussian predictions, this is simply the cumulative distribution function $P(y) = \frac{1}{2}\left[\erf\left(\frac{y-\mu}{\sqrt{2}\sigma}\right) + 1\right]$, where $y$ is the observed outcome, $\mu$ is mean of the predicted normal distribution, and $\sigma$ its standard deviation.
Once the list of all these probabilities is computed, the empirical distribution function associated to such list represents the reliability diagram. Once plotted, the range of assigned probabilities (from 0 to 1) is on the horizontal axis, and the frequency with which events occur, for each given assigned probability, is on the vertical axis.
A perfectly calibrated model results in the reliability diagram following a straight diagonal line, while over- or under-confident predictions lie respectively below or above the diagonal line.} 

In any decision-making scenario, reliability is as important as accuracy: a non-reliable model (either because over- or under-confident) introduces a systematic bias which is hard to account for. 
{In summary, reliability gives a quantitative measure of how consistently trustworthy (reliable, in common language) a predictive model is.}

\subsection{Proposed strategy}
Our method is very general and decoupled from any particular choice for the model that predicts the output targets, which can lie anywhere in the range from white to black-box models, as long as the quantity of interest is real and continuous. Indeed, in the following we will assume that such a model, whose details are not important, is provided. It is important to emphasize that the scope of this work is not to reduce the errors associated with the model, but to estimate the uncertainty of its output, thus generating a probabilistic forecast based on a deterministic model.
The probabilistic forecast is designed to be a Gaussian probability distribution centered around the values produced by the model. In this way, the only unknown quantity is the variance of the Gaussian distribution.
The simple strategy proposed here is to estimate this unknown variance (which is in general a function of the model inputs) by enforcing it to be a minimizer of a newly introduced cost function, which encodes a trade-off between accuracy and reliability, and that we call Accuracy-Reliability (AR) cost function.
As we will show, when interpreted as a function of the variance (or its square root, the standard deviation), for fixed errors (the difference between model output and observed values), accuracy and reliability are competing objectives. This gives rise to a two-objective optimization problem and the well-known Pareto curve \citep{branke08}. This curve defines a boundary on which any further optimization of one objective (e.g. a better accuracy) results in worsening of the other objective (e.g. a worse reliability).\\
{An important consideration is that, because our method boils down to a multi-dimensional optimization problem, it could in principle be directly solved using a standard algorithm such as Newton's or quasi-Newton's method. This would however result in a non-smooth function between inputs and variance, and it would not be easily generalizable to unseen inputs. Therefore, }
in order to be able to generalize the derivation of the optimal variance for any values of model inputs, and to ensure that the variance is to a certain degree a smooth function of the inputs, we introduce an Artificial Neural Network (ANN), that is trained on a given sample of model outputs, for which the ground truth is known (that is, the true output of interest, not the true variance, that remains a latent variable). 
{Hence, the method reduces to a straightforward implementation of an ANN that outputs the optimal variance that minimizes the AR. As a general strategy (and the one used in all our examples), one can use the same inputs used by the deterministic model in the neural network. However, if some prior information is known about latent variables, other inputs can be used as well.}\\
The paper is organized as follows. Section \ref{sec:methodology} introduces the mathematical background, the Accuracy-Reliability cost function, and it explains the methodology to derive the unknown uncertainties. Section \ref{sec:experiments} demonstrate the use of our methods for synthetic data, and real-world examples relevant to space weather forecasting are presented in Sections \ref{section:dend2d} and \ref{section:chorus}.
Finally, conclusions are drawn in Section \ref{sec:conclusions}.

\section{Methodology}\label{sec:methodology}
In this Section we introduce and discuss the Continuous Rank Probability Score, which is widely used in many applications \citep{matheson76}, and the new Reliability Cost for Gaussian forecasts.

\subsection{Continuous Rank Probability Score}
The Continuous Rank Probability Score (CRPS) is a generalization of the well-known Brier score \citep{wilks11}, used to assess the probabilistic forecast of continuous scalar variables, when 
the forecast is given in terms of a probability density function, or its cumulative distribution. CRPS is defined as 
\begin{linenomath}
\begin{equation}
  \text{CRPS} = \int_{-\infty}^\infty \left[P(y) - H(y-y^o) \right]^2 dy
\end{equation}
\end{linenomath}
where $P(y)$ is the cumulative distribution (cdf) of the forecast, $H(y)$ is the Heaviside function, and $y^o$ is the true (observed) value of the forecasted variable.
CRPS is a negatively oriented score: it is unbounded and equal to zero for a perfect forecast with no uncertainty (deterministic). 
In this paper we restrict our attention to the case of probabilistic forecast in the form of Gaussian distributions. Hence, a forecast is simply given by the mean value $\mu$
and the variance $\sigma^2$ of a Normal distribution. In this case $P(y) = \frac{1}{2}\left[\erf\left(\frac{y-\mu}{\sqrt{2}\sigma}\right) + 1\right]$
and the CRPS can be calculated analytically \citep{gneiting05} as
\begin{linenomath}
\begin{equation}\label{CRPS}
\text{CRPS}(\mu,\sigma,y^o) = \sigma\left[\frac{y^o-\mu}{\sigma}\erf\left(\frac{y^o-\mu}{\sqrt{2}\sigma}\right) + \sqrt{\frac{2}{\pi}}\exp\left(-\frac{(y^o-\mu)^2}{2\sigma^2} \right) -\frac{1}{\sqrt{\pi}}\right]
\end{equation}
\end{linenomath}

Several interesting properties of the CRPS have been studied in the literature. Notably, its decomposition into reliability and uncertainty has been shown in \citet{hersbach00}. The CRPS has the same unit as the variable of interest, and it collapses to the Absolute Error $|y^o-\mu|$ for $\sigma\rightarrow 0$, that is when the forecast becomes deterministic. CRPS is defined for a single instance of forecast and observation, hence it is usually averaged over an ensemble of predictions of size $N$, to obtain the score relative to a given model: $\overline{\text{CRPS}} = \sum_k \text{CRPS}(\mu_k,\sigma_k,y^o_k)/N$. 
Since we are approaching the problem of variance estimation by assigning an empirical variance to predictions originally made as single-point estimates, it makes sense to minimize the CRPS as a function of $\sigma$ only, for a fixed value of the error $\varepsilon=y^o-\mu$.
By differentiating Eq.(\ref{CRPS}) with respect to $\sigma$, one obtains
\begin{linenomath}
\begin{equation}
 \frac{d \text{CRPS}}{d\sigma} = \sqrt{\frac{2}{\pi}}\exp\left(-\frac{\varepsilon^2}{2\sigma^2} \right) -\frac{1}{\sqrt{\pi}}
\end{equation}
\end{linenomath}

and the minimizer is found to be
\begin{linenomath}
\begin{equation}\label{sigma_min}
 \sigma_{\text{min}}^{\text{CRPS}} = \frac{\varepsilon}{\sqrt{\log 2}}.
\end{equation}
\end{linenomath}

\begin{figure}[ht]
\begin{center}
\centerline{\includegraphics[width=.7\columnwidth]{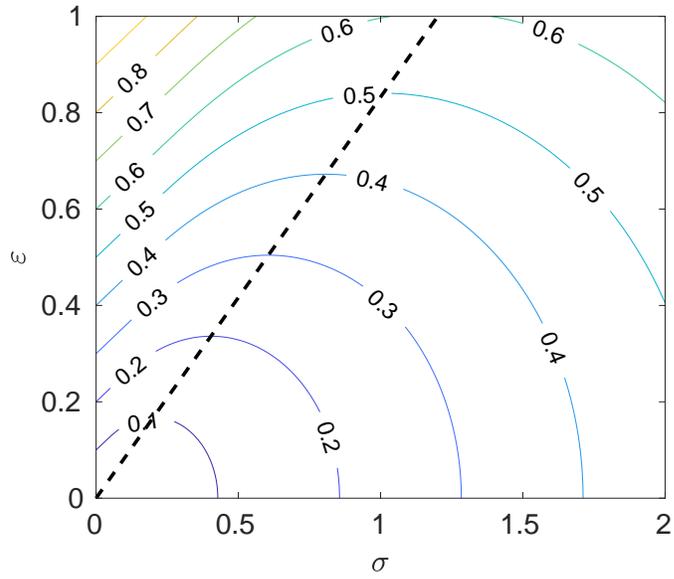}}
\caption{Lines of constant CRPS in $(\sigma,\varepsilon)$. The value of CRPS is indicated on the isolines. The black dashed line shows the location of $\sigma_{\text{min}}$ (i.e. the smallest CRPS for a given $\varepsilon$).}
\label{fig:CRPS}
\end{center}
\end{figure}

The CRPS penalizes under- and over-confident predictions in a non-trivial way. Indeed, for any value of the error $\varepsilon$, there are always two values of $\sigma$ (one smaller and one larger than $\sigma_{\text{min}}$, that is one over- and the other under-confident) that yield the same CRPS. We show in Figure \ref{fig:CRPS} the isolines of CRPS in $(\sigma,\varepsilon)$ space. The black dashed line indicates $\sigma_{\text{min}}$. From this Figure it is clear how a smaller error $\varepsilon$ (for constant $\sigma$) always results in a smaller (better) score, but the same score can be achieved by changing both the error $\varepsilon$ and the standard deviation $\sigma$. A straightforward way of understanding how CRPS works is the following.
Let us start with a prediction that has a given error $\varepsilon$ and no uncertainty (i.e. a deterministic forecast, $\sigma=0$). CRPS attributes a certain score to such prediction. Now, if we increase $\varepsilon$ the prediction becomes obviously worse, hence CRPS increases, unless we simultaneously increase the uncertainty $\sigma$. That is, accounting for the fact that the prediction is uncertain compensates for a larger error. In this way one can move along a constant CRPS curve, until the point (on the dashed line) where an increase in error cannot be compensated any further by an increase in uncertainty. After that point, larger uncertainties must then be compensated by a \emph{decrease} in the error $\varepsilon$.

\subsection{Reliability Score for Gaussian forecast}
Contrary to the CRPS, that is defined for a single pair of forecast-observation, it is clear that reliability can only be defined for a large enough ensemble of such pairs, being a statistical property of a model. 
In the case of normally distributed forecasts, we expect {the standardized errors $\eta$ defined as 
\begin{equation}
\eta=\varepsilon/(\sqrt{2}\sigma)
\end{equation}
}
calculated over a sample of $N$ predictions-observations to have a standard normal distribution with cdf $\Phi(\eta)=\frac{1}{2}(\erf(\eta)+1)$. Hence, we define the Reliability Score (RS) as:
\begin{linenomath}
\begin{equation}\label{RS_1}
 \text{RS} = \int_{-\infty}^\infty \left[\Phi(y) - C_\eta(y)\right]^2 dy
\end{equation}
\end{linenomath}

where $C_\eta(y)$ is the empirical cumulative distribution of the {standardized} errors $\eta$, that is 
\begin{linenomath}
\begin{equation}
 C_\eta(y) = \frac{1}{N}\sum_{i=1}^N H(y-\eta_i)
\end{equation}
\end{linenomath}

with $\eta_i = (y^o_i-\mu_i)/(\sqrt{2}\sigma_i)$. RS measures the divergence of the empirical distribution of {standardized} errors $\eta$ from a standard normal distribution. 
{Note that, by appropriately choosing $\sigma$, one can always obtain an approximate standard normal distribution for $\eta$, for any given distribution of the errors $\varepsilon=y^o-\mu$, as long as the number of instances for $\varepsilon<0$ and $\varepsilon>0$ are approximately equal}.
From now on we will use the convention that the set $\eta=\{\eta_1,\eta_2,\ldots \eta_N\}$ is sorted ($\eta_i\leq\eta_{i+1}$). This does not imply that $\mu_i$ or $\sigma_i$ are sorted as well.
Interestingly, the integral in Eq. (\ref{RS_1}) can be calculated analytically, via expansion into a telescopic series, yielding:
\begin{linenomath}
\begin{equation}\label{RS}
  \text{RS} = \sum_{i=1}^N \left[\frac{\eta_i}{N}\left(\erf(\eta_i)+1\right) - \frac{\eta_i}{N^2}(2i-1) + \frac{\exp(-\eta_i^2)}{\sqrt{\pi}N}\right] -\frac{1}{2}\sqrt{\frac{2}{\pi}}
\end{equation}
\end{linenomath}

\normalsize
Differentiating now the $i$-th term of the above summation, RS$_i$, with respect to $\sigma_i$ (for fixed $\varepsilon_i$), one obtains
\begin{linenomath}
\begin{equation}
 \frac{d\text{RS}_i}{d\sigma_i} =   \frac{\eta_i}{N\sigma_i}\left(\frac{2i-1}{N}-\erf(\eta_i)-1 \right),
\end{equation}
\end{linenomath}

which is minimized at the value $\sigma_{\text{min}}^{\text{RS}}$ that satisfies
\begin{linenomath}
\begin{equation}\label{optimal_eta}
 \erf\left(\frac{\varepsilon_i}{\sqrt{2}\sigma_{\text{min}}^{\text{RS}}}\right) = \frac{2i-1}{N}-1.
\end{equation}
\end{linenomath}
This could have been trivially derived by realizing that by minimizing RS one obtains the distribution of {standardized} errors $\eta_i$ that most closely approximates a standard normal distribution, for a given number of observations $N$. This is the distribution that mapped through $\Phi$ divides uniformly the interval $[0,1]$:
$\frac{1}{2}(\erf(\eta_i)+1)=\frac{i-1/2}{N}$, i.e. the set $\{\frac{1}{2N},\frac{1}{2N}+\frac{1}{N},\frac{1}{2N}+\frac{2}{N},\ldots,1-\frac{1}{2N}\}$. Like CRPS, RS is negatively oriented {(i.e. zero is the perfect score)}. It can be equal to zero only for $N\rightarrow \infty$.

\subsection{The Accuracy-Reliability cost function}
The Accuracy-Reliability cost function introduced here follows from the simple principle that the empirical variances $\sigma_i^2$ estimated from an ensemble of errors $\varepsilon_i$ should result in a model that is both accurate (with respect to the CRPS score), and reliable (with respect to the RS score). This gives rise to a two-objective optimization problem. It is trivial to verify that CRPS and RS cannot simultaneously attain their minimum value (for fixed errors $\varepsilon_i$). Note that CRPS is a function of $\varepsilon_i$ and $\sigma_i$, while RS is only a function of their scaled ratio $\eta_i=\varepsilon_i/(\sqrt{2}\sigma_i)$. By minimizing the CRPS, $\eta_i = \frac{1}{2}\sqrt{\log 4}$ for any $i$ (see Eq. \ref{sigma_min}). Obviously, a constant $\eta_i$ cannot result in a minimum also for RS, according to Eq. (\ref{optimal_eta}). Moreover, notice that trying to minimize RS as a function of $\sigma_i$ (for fixed errors $\varepsilon_i$) result in an ill-posed problem, because one can have infinite combinations of $\sigma_i$ that result in the same set $\eta$, therefore there is no unique solution for the variances that minimizes RS. Hence, RS can be thought of as a regularization term in the Accuracy-Reliability cost function. The simplest strategy to deal with multi-objective optimization problems is to scalarize the cost function, which we define here as
\begin{linenomath}
\begin{equation}\label{AR}
 \text{AR} = \beta\cdot \overline{\text{CRPS}} + (1-\beta)\text{RS}.
\end{equation}
\end{linenomath}

We choose the scaling factor $\beta$ as
\begin{linenomath}
\begin{equation}\label{beta}
 \beta={\text{RS}}_{min}/(\overline{\text{CRPS}}_{min} + \text{RS}_{min}).
\end{equation}
\end{linenomath}

The minimum of $\overline{\text{CRPS}}$ is $\overline{\text{CRPS}}_{min}=\frac{\sqrt{\log 4}}{2N}\sum_{i=1}^N \varepsilon_i$, which is simply the mean of the errors, rescaled by a constant. 
The minimum of RS follows from Eqs. (\ref{RS}) and (\ref{optimal_eta}): 
\begin{linenomath}
\begin{equation}
\text{RS}_{min} = \frac{1}{\sqrt{\pi} N}\sum_{i=1}^N \exp\left(-\left[\erf^{-1}\left(\frac{2i-1}{N}-1\right)\right]^2\right)-\frac{1}{2}\sqrt{\frac{2}{\pi}}
\end{equation}
\end{linenomath}
Notice that $\text{RS}_{min}$ is only a function of the size of the sample $N$, and it converges to zero for $N\rightarrow \infty$. 
The heuristic choice in Eq. (\ref{beta}) is justified by the fact that the two scores might have different orders of magnitude, and therefore we rescale them in such a way that they are comparable in our cost function (\ref{AR}). Indeed the scaling factor $\beta$ ensure that the two terms would be exactly equal if both could be minimized simultaneously.
We believe this to be a sensible choice, although there might be applications where one would like to weigh the two scores differently.  Also, in our practical implementation, we neglect the last constant term in the definition (\ref{RS}) so that, for sufficiently large $N$, $\text{RS}_{min}\simeq \frac{1}{2}\sqrt{\frac{2}{\pi}}\simeq 0.4$

\subsection{Neural Network}

In summary, we want to estimate the input-dependent values of the empirical variances $\sigma_i^2$ associated to a sample of $N$ observations for which we know the errors $\varepsilon_i$. We do so by solving a multidimensional optimization problem in which the set of estimated $\sigma_i$ minimizes the AR cost function defined in Eq. (\ref{AR}). This newly introduced cost function has a straightforward interpretation as the trade-off between accuracy and reliability, which are two essential but conflicting properties. In practice, we want to generate a model that is able to predict $\sigma^2$ as a function of the inputs $\mathbf{x}$ on any point of a domain. {This unknown function can in general be non-linear, and we assume no a-priori information to constraint its functional form. However, we want to enforce smoothness of the unknown variance, to some degree. A very general strategy is to use a regularized Artificial Neural Network to model the dependency of $\sigma^2$ as a function of the inputs. However, it is important to realize that this is not the only choice, and in case the user has some prior information on the functional form of $\sigma^2$, other strategies (such as polynomial interpolation, if the input is low-dimensional) might be better suited.}
For simplicity, we choose a single neural network architecture, that we use for all the tests. We use a network with 2 hidden layers, respectively with 20 and 5 neurons. The activation functions are $\tanh$ and a symmetric saturating linear function, respectively. The third (output) layer uses a linear activation function. The dataset, composed of the inputs $\mathbf{x}$ and the corresponding observed errors $\varepsilon$, is randomly divided into training ($70\%$) and validation ($30\%$) sets. The network is trained using a standard Broyden-Fletcher-Goldfarb-Shanno (BFGS) quasi-Newton algorithm, and the iterations are forcefully stopped when the loss function does not decrease for 10 successive iterations on the validation set. These are all standard choices when training neural networks, and we refer the reader to specific monographs \citep[e.g.,][]{bishop95}.
A very attractive feature of our model is that the only inputs needed are the input parameters $x_i$ and the corresponding errors $\varepsilon_i$ {(used for training only)}. The neural network outputs the values of $\log(\sigma_i)$, by minimizing the above-introduced Accuracy-Reliability cost function, Eq. (\ref{AR}), where $\sigma$ is the standard deviation, and $\log$ is used to enforce its positivity. In order to limit the expressive power and avoid over-fitting, we may add a regularization term equal to the L$_2$ norm of the weights to the AR cost function, {multiplied by a constant factor 0.2. In other words, a term $0.2\frac{w^Tw}{2}$ can be added to the AR cost function defined in Eq. (\ref{AR}), where the vector $w$ represents the Neural Network weights. This is a standard procedure to constrain the amplitude of the weights, and avoid over-fitting (because highly nonlinear functions tend to increase the regularization term) \citep[see, e.g][]{care18}. In our numerical experiments (Section \ref{sec:experiments}) this regularization term was needed only for 1D cases.} Finally, in order to avoid local minima due to the random initialization of the neural network weights, we train five independent networks and choose the one that yields the smallest value of the cost function.

\section{Experiments with synthetic data}\label{sec:experiments}
\begin{figure}
\begin{center}
\centerline{\includegraphics[width=.5\columnwidth]{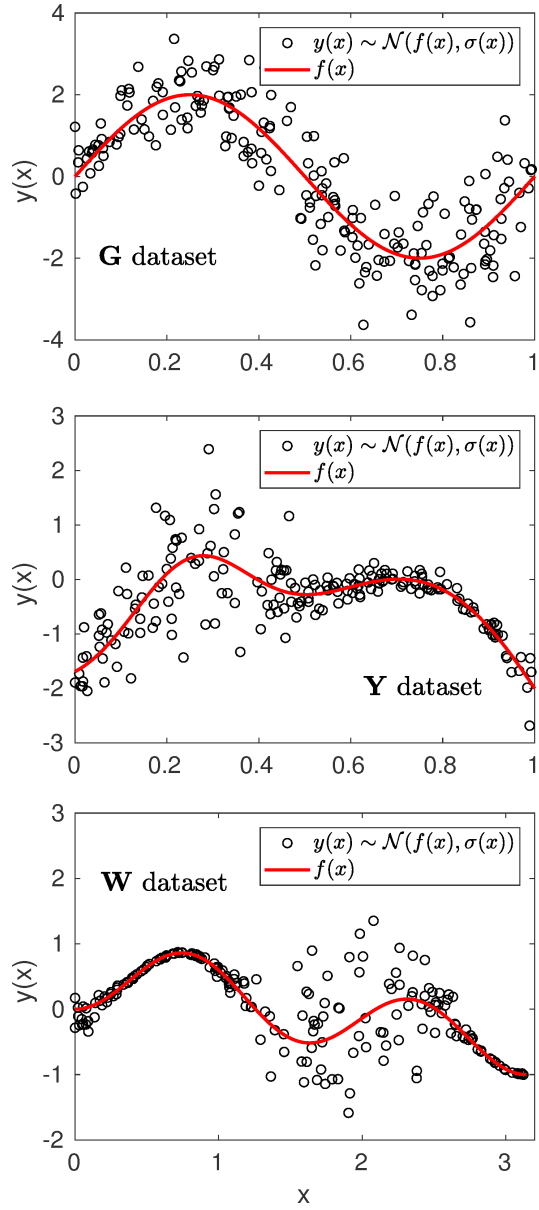}}
\caption{Circles: 200 points sampled from the {\bf G}, {\bf Y}, {\bf W} dataset (top, middle, bottom, respectively). The red line shows the mean function $f(x)$.}
\label{G_regression}
\end{center}
\end{figure}

\begin{figure}[ht]
\begin{center}
\centerline{\includegraphics[width=.7\columnwidth]{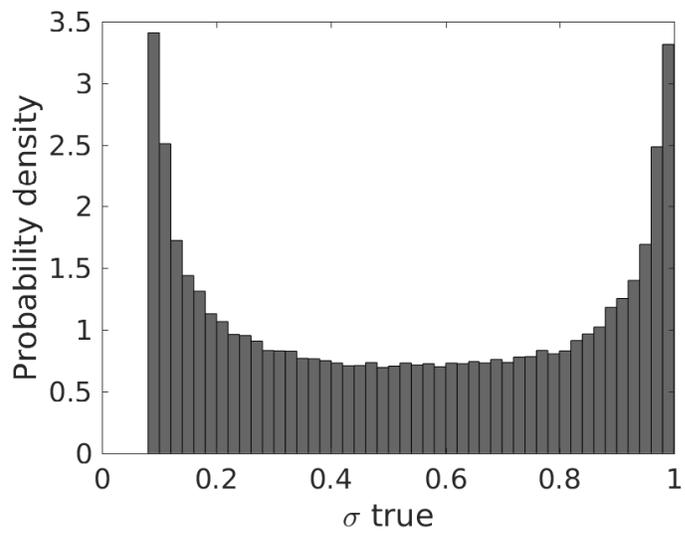}}
\caption{Distribution of true values of standard deviation $\sigma$ for the {\bf 5D} dataset.}
\label{multiD_2}
\end{center}
\end{figure}

\begin{figure}[ht]
\begin{center}
\centerline{\includegraphics[width=.5\columnwidth]{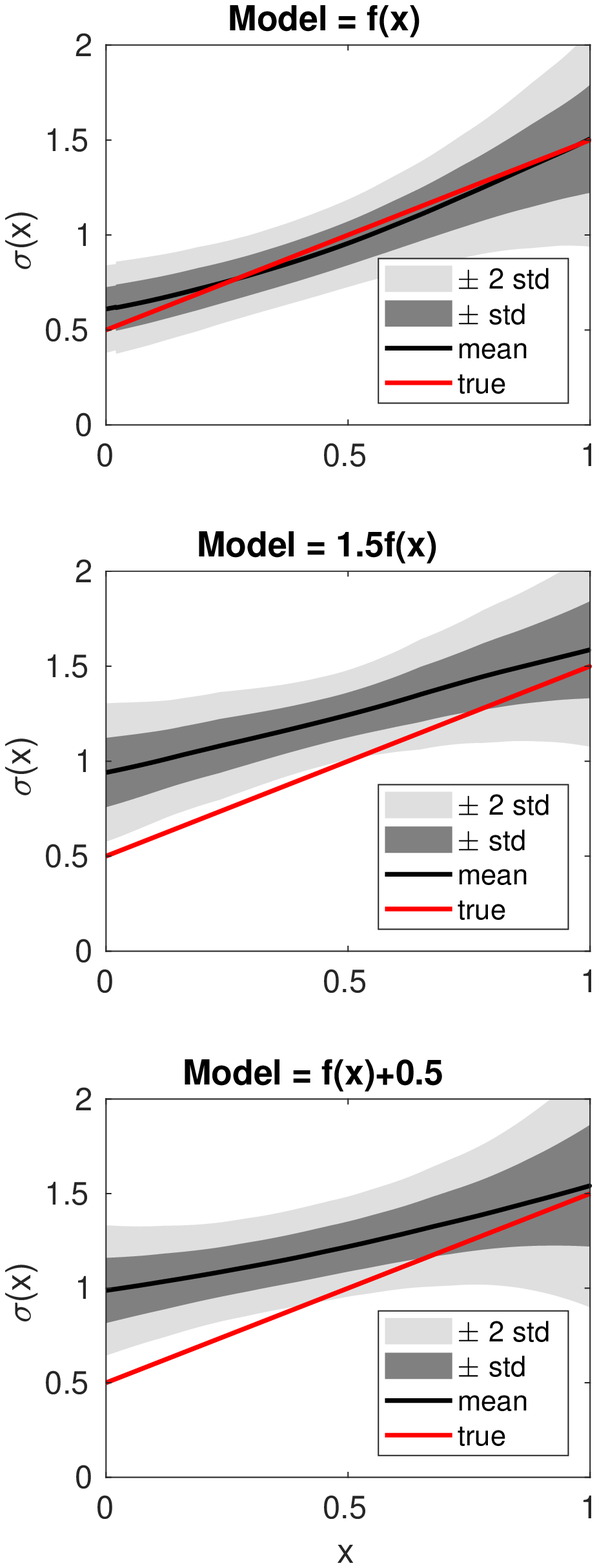}}
\caption{Results for the {\bf G} dataset. Values derived for the standard deviation $\sigma$, averaged over 200 independent runs (black), compared to the ground truth values used to generate the data (in red). The shaded gray area represents the confidence intervals of one (dark gray) and two (light gray) standard deviations calculated over the ensemble of 200 runs. Top: the correct mean function $f(x)$ is used for the model; middle: a mis-specified model that uses $1.5f(x)$ as mean; bottom: a mis-specified model that uses $f(x)+0.5$ as mean.}
\label{fig:G_results}
\end{center}
\end{figure}

\begin{figure}[ht]
\begin{center}
\centerline{\includegraphics[width=.7\columnwidth]{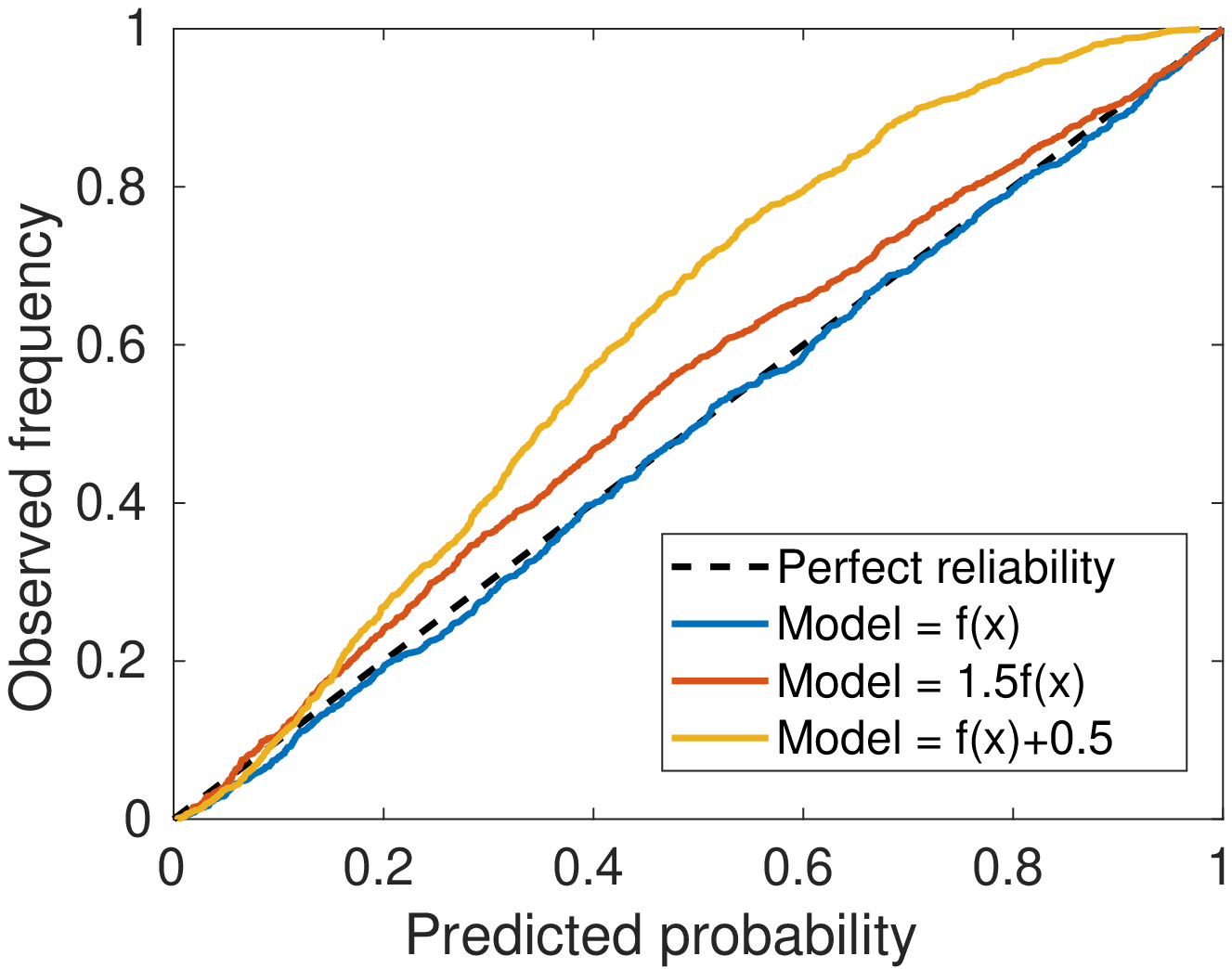}}
\caption{Reliability diagram for the method applied to the {\bf G} dataset. Blue, red and yellow lines denote the observed frequency as function of the predicted probability, for the cases of correct mean function $f(x)$, and mis-specified models $1.5f(x)$ and $f(x)+0.5$, respectively. A perfect reliability is shown as a black dashed line.}
\label{fig:G_rel}
\end{center}
\end{figure}

In this section we show some experiments on synthetic data, to demonstrate the ease, robustness and accuracy of the presented method to derive uncertainties.
Here, we assume to have an imperfect model that produces a forecast $y=f(x)$.  
The synthetic observations are generated from a Gaussian distribution $\mathcal{N}(f(x),\sigma(x)^2)$, with known variance $\sigma(x)^2$. The stochastic nature of the synthetic data can be thought to mimic the existence of latent variables that are not included in the model. In other words, close values of the input $x$ can results in very different outputs, because of unmodeled processes. The purpose of these experiments is to show that our method is capable of recovering the functional dependence of the variance $\sigma(x)^2$, that is, for real data, unknown.
We choose some of the datasets routinely used in machine learning literature \citep{kersting07}. The first three datasets are one-dimensional in $x$, while in the fourth we will test the method on a five-dimensional space, thus showing the robustness of the proposed strategy.\\
{\bf G} dataset: $x \in [0,1]$, $f(x) = 2\sin(2\pi x)$, $\sigma(x) = x+\frac{1}{2}$ \citep{goldberg98}.\\ 
{\bf Y} dataset: $x \in [0,1]$, $f(x) = 2(\exp(-30(x-0.25)^2)+\sin(\pi x^2))-2$, $\sigma(x) = \exp(\sin(2\pi x))$ \citep{yuan04}. \\
{\bf W} dataset: $x \in [0,\pi]$, $f(x) = \sin(2.5x)\sin(1.5x)$, $\sigma(x) = 0.01+0.25(1-\sin(2.5x))^2$ \citep{weigend94, williams96}. \\
Examples of 200 points sampled from the {\bf G}, {\bf Y} and {\bf W}  dataset are shown in Figure \ref{G_regression} along with their mean function $f(x)$ in red.\\
For the {\bf G}, {\bf Y}, and {\bf W} datasets we test the case where the true mean function $f(x)$ is used as deterministic model, and two cases where the model suffers of a systematic bias and the model output is replaced by $\frac{3}{2}f(x)$ (a multiplicative error) or $f(x)+\frac{1}{2}$ (an additive error). These two cases serve also the purpose of studying the behaviour of the proposed method for non-Gaussian errors. Every model is trained on 100 points uniformly sampled in the domain.\\ 
{\bf 5D} dataset: $\mathbf{x} \in [0, 1]^5$, $f(\mathbf{x})=0$, $\sigma(\mathbf{x})=0.45(\cos(\pi + \sum_{i=1}^5 5x_i) + 1.2)$ \citep{genz84}.
Figure \ref{multiD_2} shows the distribution of $\sigma$, which ranges in the interval $[0.09,0.99]$.\\
The {\bf 5D} dataset is obviously more challenging, hence we use 10,000 points to train the model (note that this results in fewer points per dimension, compared to the one-dimensional tests).
For all experiments we test 200 independent runs.\\
The results for the {\bf G} dataset are shown in Figure \ref{fig:G_results}. The values derived for the standard deviation $\sigma$, averaged over 200 independent runs are shown in black, compared to the ground truth value used to generate the data (in red). The shaded gray area represents the confidence intervals of one (dark gray) and two (light gray) standard deviations calculated over the ensemble of 200 runs. The top, middle, and bottom panel show the results when the model uses the exact mean function used to generate the data $f(x)$, and when the model is mis-specified by a multiplicative error ($1.5f(x)$), or an additive error ($f(x)+0.5$), respectively. One can notice that our method is capable of recovering almost exactly the true variance (top), when the model is accurate. On the other hand, when the model is mis-specified (and the errors become non-Gaussian) the method appropriately assigns a larger uncertainty (middle and bottom panels). In particular, it is interesting that the discrepancy between the true variance and the one derived by this method is larger when the true variance is small. This is because in the regions with small (true) variance a mis-specified model (mean function) causes a larger departure from Gaussianity. Since the method is designed to assign anyway a Gaussian probability density, it necessarily results in a larger uncertainty. Nevertheless, using the AR cost function as criterion to derive the empirical variance will always results in an optimally calibrated model, meaning that ill-calibrated results are very unlikely, unless the underlying mean function is very off from the appropriate value. Figure \ref{fig:G_rel} shows the reliability diagram for the three cases discussed (exact model and mis-specified models). Once again, a reliability diagram represents, for any value of probability predicted for a given output, what is the actual observed frequency for that output (calculated on a large sample). A perfectly calibrated model results in a reliability diagram  following the straight diagonal line (dashed black).

\begin{figure}[ht]
\begin{center}
\centerline{\includegraphics[width=.5\columnwidth]{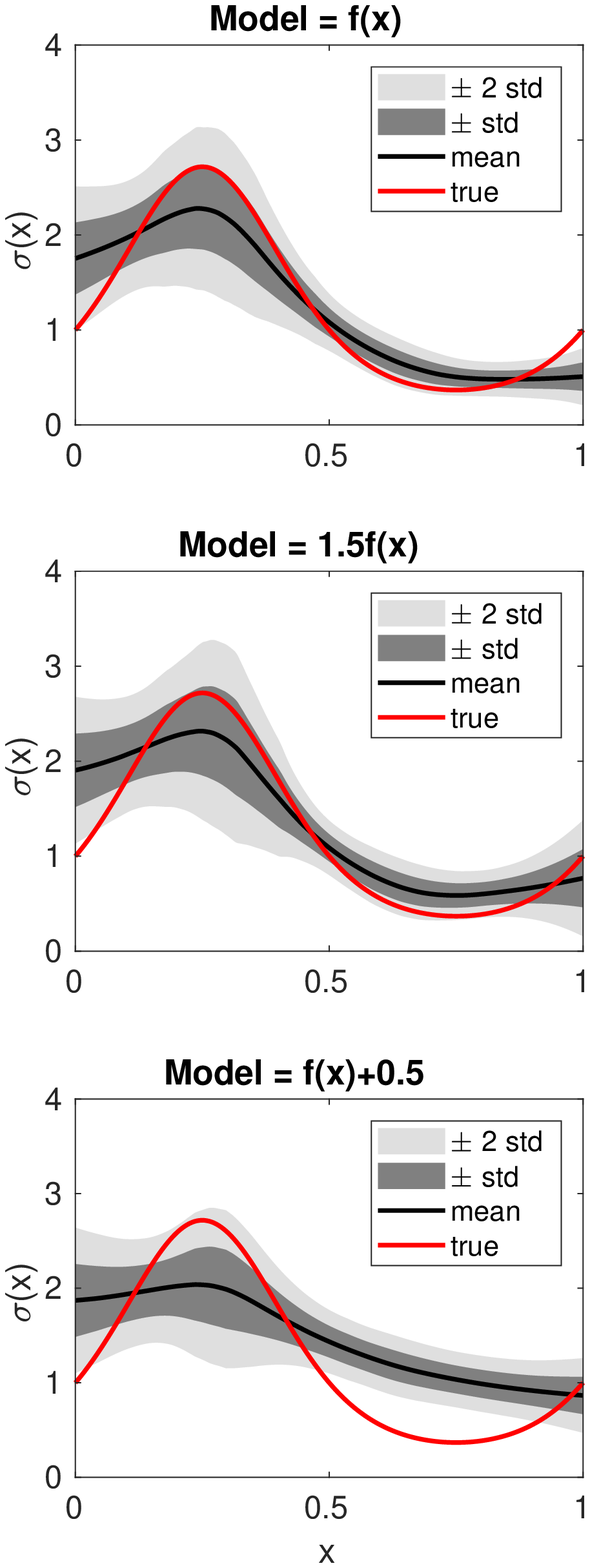}}
\caption{Results for the {\bf Y} dataset. Values derived for the standard deviation $\sigma$, averaged over 200 independent runs (black), compared to the ground truth values used to generate the data (in red). The shaded gray area represents the confidence intervals of one (dark gray) and two (light gray) standard deviations calculated over the ensemble of 200 runs. Top: the correct mean function $f(x)$ is used for the model; middle: a mis-specified model that uses $1.5f(x)$ as mean; bottom: a mis-specified model that uses $f(x)+0.5$ as mean.}
\label{fig:Y_results}
\end{center}
\end{figure}

\begin{figure}[ht]
\begin{center}
\centerline{\includegraphics[width=.7\columnwidth]{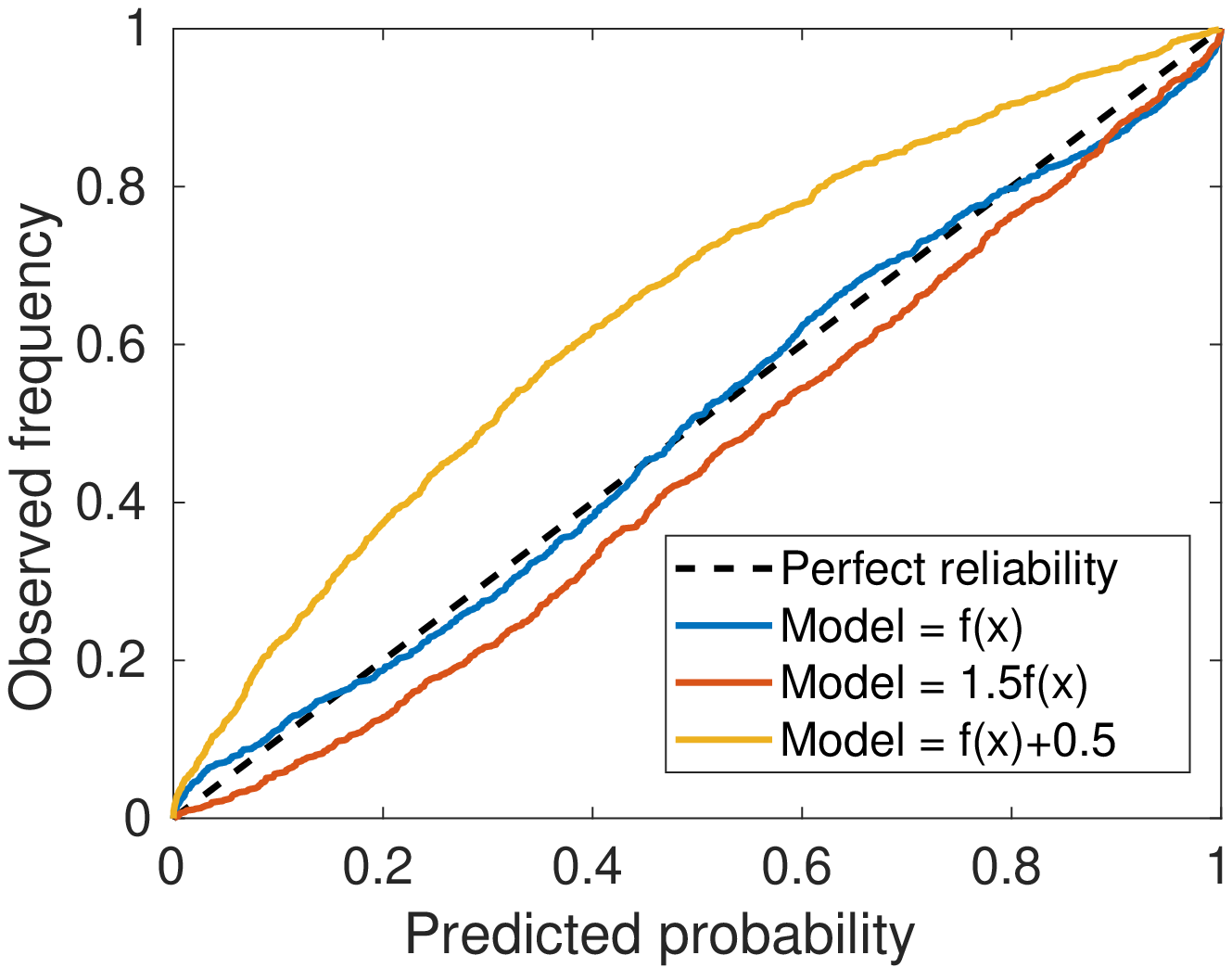}}
\caption{Reliability diagram for the method applied to the {\bf Y} dataset. Blue, red and yellow lines denote the observed frequency as function of the predicted probability, for the cases of correct mean function $f(x)$, and mis-specified models $1.5f(x)$ and $f(x)+0.5$, respectively. A perfect reliability is shown as a black dashed line.}
\label{fig:Y_rel}
\end{center}
\end{figure}

\begin{figure}[ht]
\begin{center}
\centerline{\includegraphics[width=.5\columnwidth]{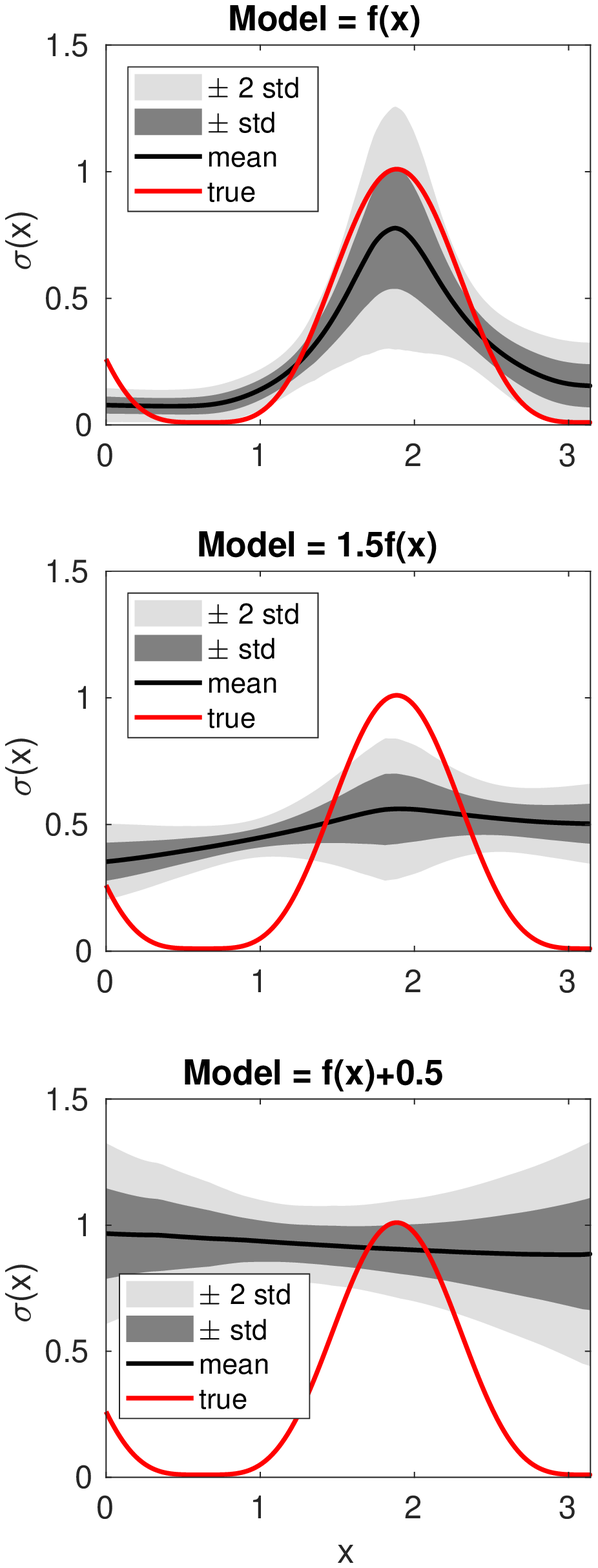}}
\caption{Results for the {\bf W} dataset. Values derived for the standard deviation $\sigma$, averaged over 200 independent runs (black), compared to the ground truth values used to generate the data (in red). The shaded gray area represents the confidence intervals of one (dark gray) and two (light gray) standard deviations calculated over the ensemble of 200 runs. Top: the correct mean function $f(x)$ is used for the model; middle: a mis-specified model that uses $1.5f(x)$ as mean; bottom: a mis-specified model that uses $f(x)+0.5$ as mean.}
\label{fig:W_results}
\end{center}
\end{figure}

\begin{figure}[ht]
\begin{center}
\centerline{\includegraphics[width=.7\columnwidth]{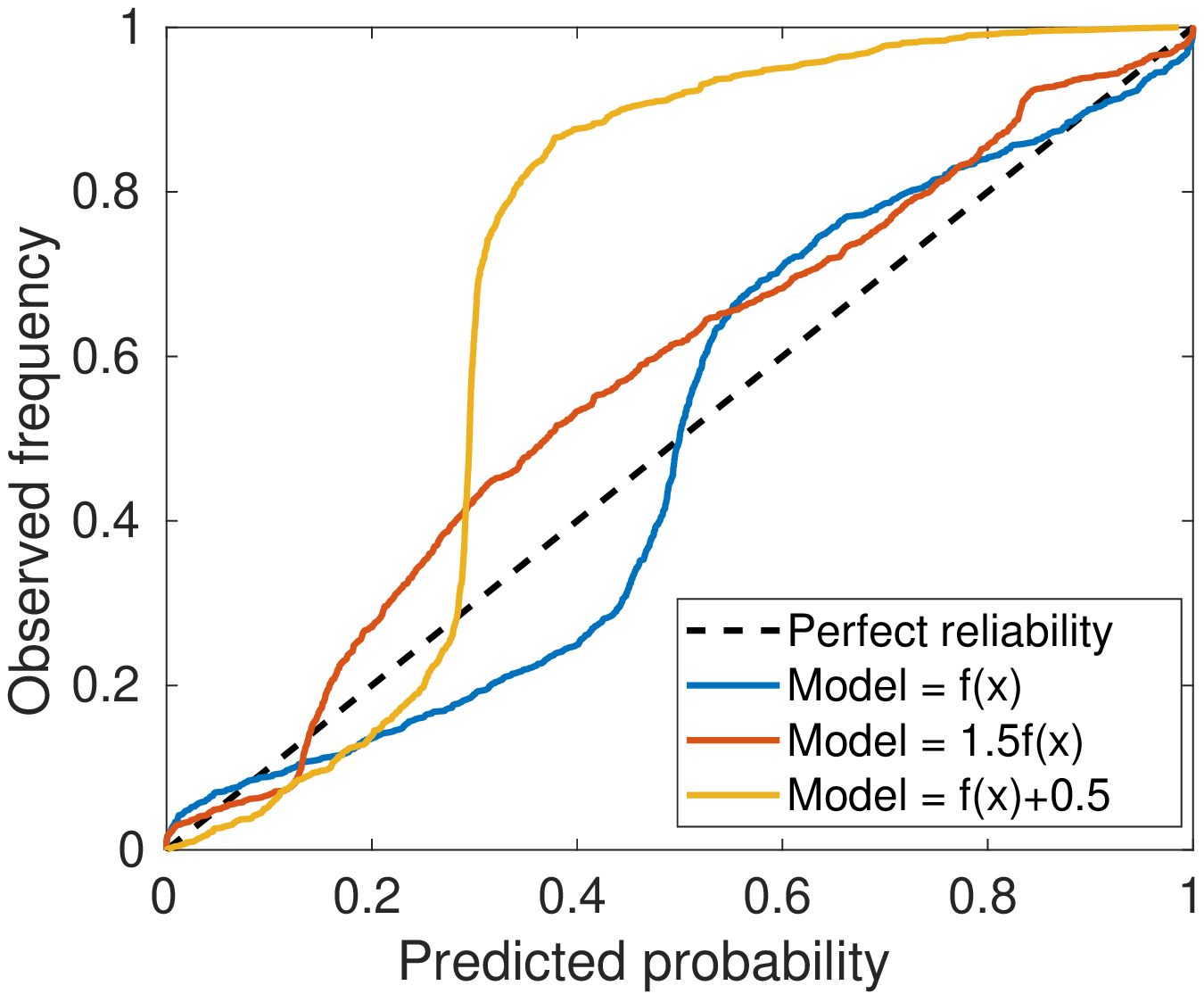}}
\caption{Reliability diagram for the method applied to the {\bf W} dataset. Blue, red and yellow lines denote the observed frequency as function of the predicted probability, for the cases of correct mean function $f(x)$, and mis-specified models $1.5f(x)$ and $f(x)+0.5$, respectively. A perfect reliability is shown as a black dashed line.}
\label{fig:W_rel}
\end{center}
\end{figure}

\begin{figure}[ht]
\begin{center}
\centerline{\includegraphics[width=.7\columnwidth]{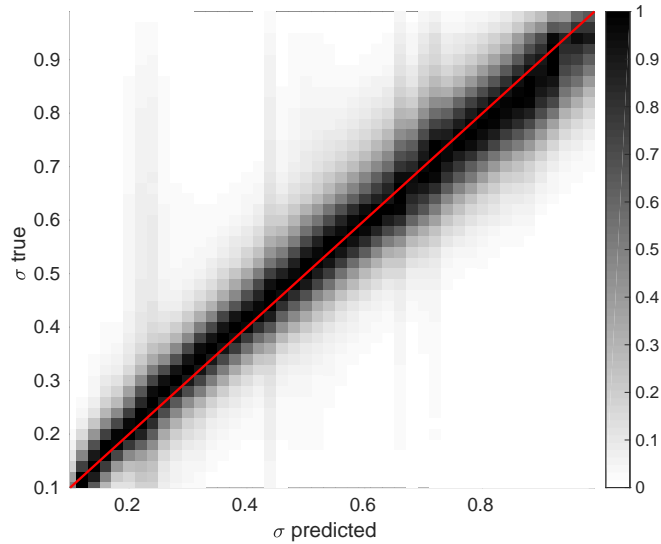}}
\caption{Probability density of the prediction versus real values of $\sigma$ for the {\bf 5D} dataset. The red line denotes perfect prediction. The densities are normalized to have maximum value along each column equal to one. 10,000,000 samples have been used to generate the plot (with a training set of 10,000 points).}
\label{multiD_1}
\end{center}
\end{figure}

Not surprisingly, when we use the exact model as our mean function (blue line), the empirical variance derived with our method result in a perfectly calibrated model that indeed follow very closely the diagonal line (dashed black).
When the model is mis-specified (red and yellow lines), the method tries to achieve a trade-off between reliability and accuracy. The resulted reliability is still very good even though not perfect. It is very interesting that the reliability diagram can be used for our method to detect a mis-specified mean function. Indeed, it is important to point out that, for the {\bf G} dataset, the model with additive error is worse than the one with multiplicative error, because $f(x)$ goes through zero in three points in the domain (hence the multiplicative error plays no role in those points). \\
Results for the {\bf Y} datasets are shown in Figures \ref{fig:Y_results} and \ref{fig:Y_rel}, with same format as previous Figures. Conclusions are very similar, with the main difference that the {\bf Y} dataset has a nonlinear true variance, which is harder to learn. Nevertheless, our method provides a good estimate of it.
The {\bf W} model is the most challenging, as shown in Figures \ref{fig:W_results}, \ref{fig:W_rel}.  Here, a mis-specification of the model becomes readily evident, producing almost constant variance and large errors in the reliability diagram.\\
For the {\bf 5D} dataset it is impractical to compare graphically the real and estimated $\sigma(\mathbf{x})$ in the 5-dimensional domain. Instead, in Figure \ref{multiD_1} we show the probability density of the real versus predicted values of the standard deviation. Values are normalized such that the maximum value in the colormap for any value of predicted $\sigma$ is equal to one (i.e. along vertical lines). The red line shows a perfect prediction. The colormap has been generated by 10,000,000 points, while the model has been trained with 10,000 points only.
For this case, we have used an exact mean function (equal to zero), in order to focus exclusively on the estimation of the variance.
We believe that this is an excellent result for a very challenging task, given the sparsity of the training set, that shows the robustness of the method.

\section{Estimation of electron density in the plasmasphere (DEN2D)}\label{section:dend2d}
\begin{figure}[ht]
\begin{center}
\centerline{\includegraphics[width=.7\columnwidth]{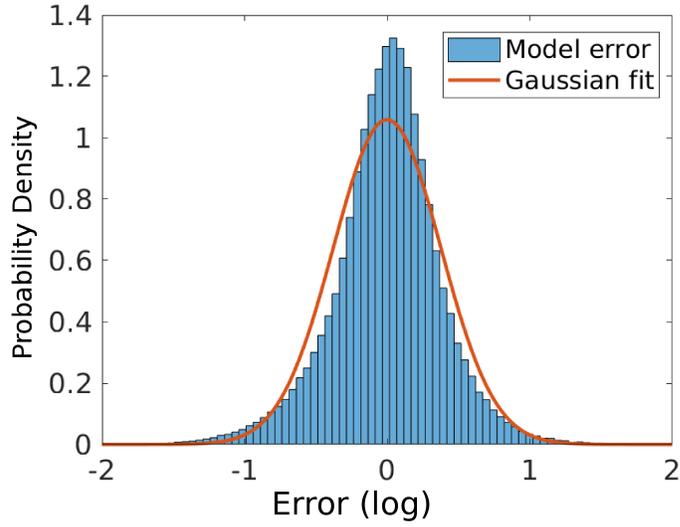}}
\caption{Density histogram of the DEN2D model errors (in logarithmic scale). The red line indicates a Gaussian fit.}
\label{fig:hist_den2d}
\end{center}
\end{figure}

\begin{figure}[ht]
\begin{center}
\centerline{\includegraphics[width=.7\columnwidth]{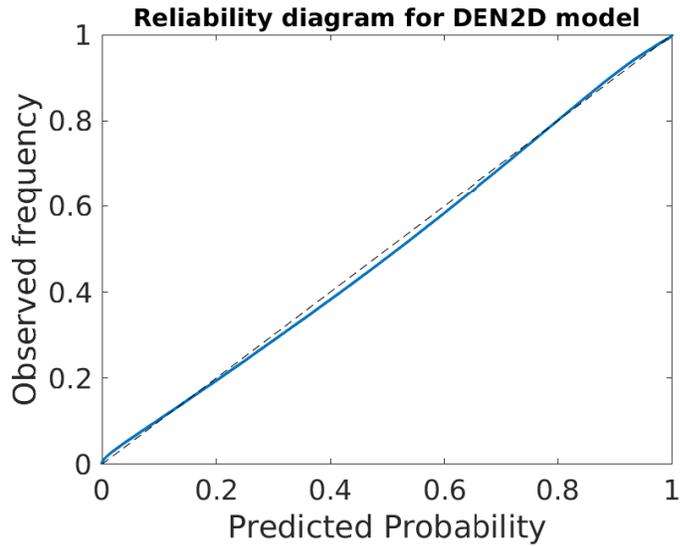}}
\caption{Reliability diagram of the probabilistic estimate of electron density, using the DEN2D model as mean function. The black dashed line indicates perfect reliability.}
\label{fig:rel_den2d}
\end{center}
\end{figure}

\begin{figure}[ht]
\begin{center}
\centerline{\includegraphics[width=.7\columnwidth]{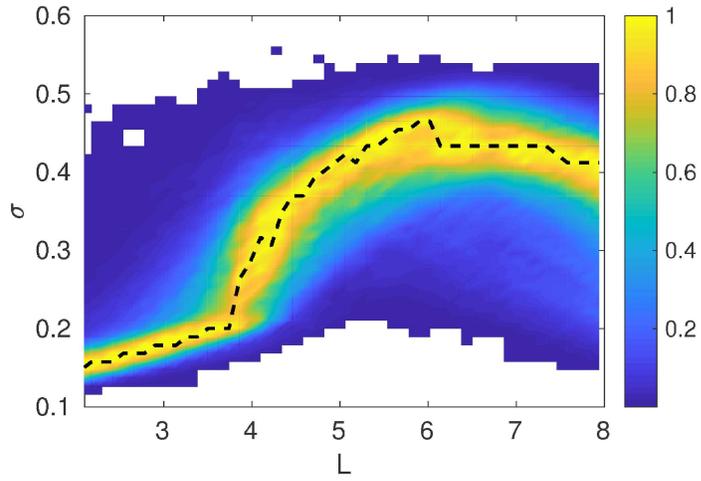}}
\caption{DEN2D model. Two-dimensional histogram of standard deviation $sigma$ versus $L$. The number of counts are normalized column-wise: the maximum for each value of $L$ is equal to 1. The black-dashed line follows the peak of the distribution as function of $L$.}
\label{fig:dist_den2d}
\end{center}
\end{figure}

\begin{figure}[ht]
\begin{center}
\centerline{\includegraphics[width=.7\columnwidth]{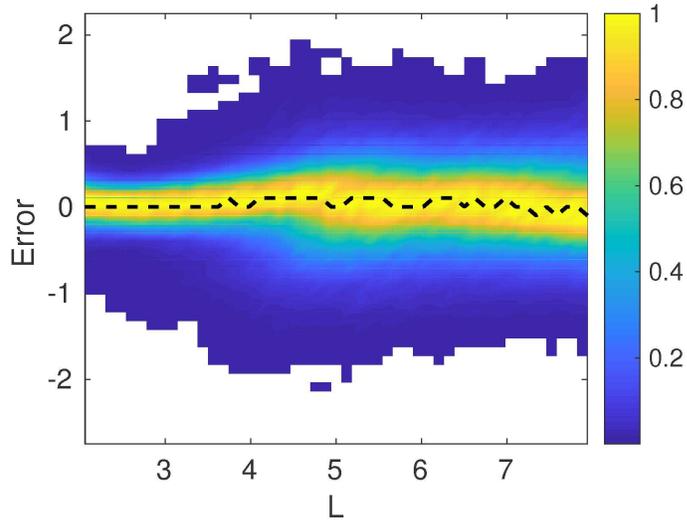}}
\caption{DEN2D model. Two-dimensional histogram of errors versus $L$. The number of counts are normalized column-wise: the maximum for each value of $L$ is equal to 1. The black-dashed line follows the peak of the distribution as function of $L$.}
\label{fig:err_den2d}
\end{center}
\end{figure}

\begin{figure}[ht]
\begin{center}
\centerline{\includegraphics[width=\columnwidth]{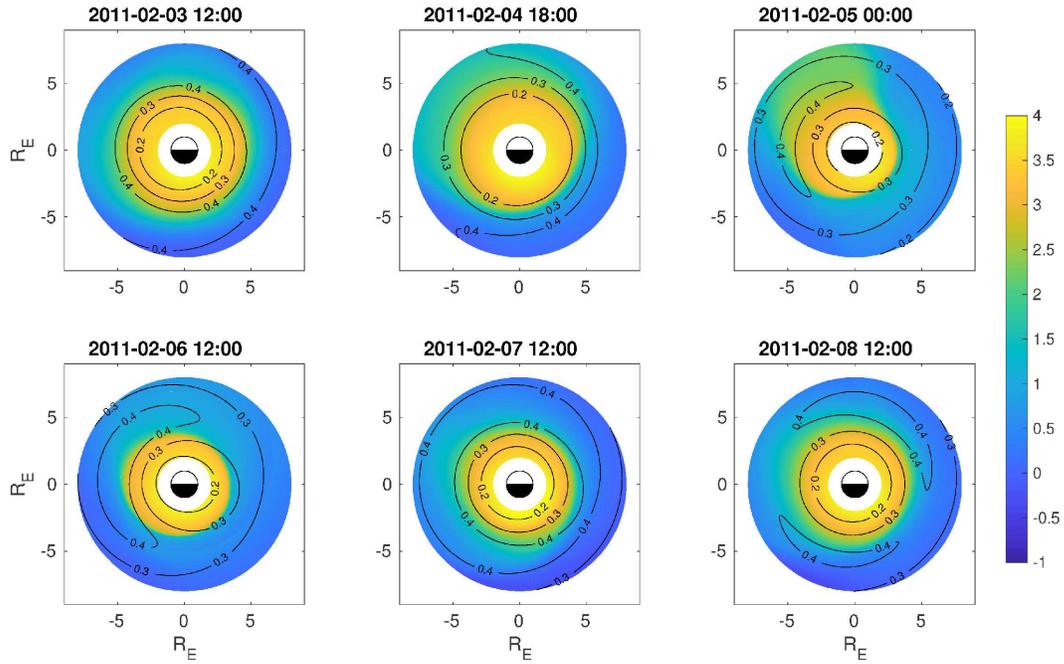}}
\caption{DEN2D model. A series of panels showing the estimated electron density (in color), and the associated standard deviation $\sigma$ (as isolines) for the event of 4 February 2011, as function of $L$ and MLT. The heat map represents the logarithm of the electron number density in el/cc (see Figure 6 in \citet{chu17}).}
\label{fig:storm_den2d}
\end{center}
\end{figure}

In this and the next section we show applications of our method that are relevant to Space Weather.
The first example is the estimation of the electron plasma density in the plasmasphere. \cite{chu17} have devised a neural network (NN) model, DEN2D, that takes as inputs the time history of the SYM-H and AL geomagnetic indexes, and of $F_{10.7}$ (solar radio flux), and outputs the logarithm of the electron density at any location in the plasmasphere, as function of magnetic shell (L), and magnetic local time (MLT), at near-equatorial latitudes.  The neural network was trained and tested on about 400,000 events generated by 4 years of THEMIS data (June 2008 to December 2012). It uses 178 input attributes and outputs the logarithmic value of the electron density.
Obviously, the NN is a deterministic model, that outputs a single value for any given combinations of inputs. Hence, this model is very well-posed for the method introduced in this paper.
Moreover, a recent study performed to evaluate the propagation of uncertainties in radiation belt ensemble simulations, has shown that the uncertainty in the electron density estimation carries most of the variance of the predicted electron fluxes \citep{camporeale16}. Therefore, the reduction of the uncertainty for electron density is a necessary step for developing reliable forecasts of electron fluxes.
Figure \ref{fig:hist_den2d} shows the distribution of the error of the NN output with respect to the true {(log)} electron density, calculated over the whole dataset. The superimposed red line shows a Gaussian fit to the distribution, which has a slighter larger variance. It is important, however, to keep in mind that our method does not assume that the model errors are normally distributed. Indeed, the method will try to enforce that the {standardized} errors $\eta$ are Gaussian, which can be achieved even for non-Gaussian errors $\varepsilon$. This is well demonstrated by the reliability diagram, which is shown in Figure \ref{fig:rel_den2d}. Our method applied to the DEN2D model produces a probabilistic estimate of electron density that has almost perfect reliability.\\
Once we have trained our model to estimate the standard deviation $\sigma$ as function of the same inputs used in the DEN2D NN, one can seek for evident relationship between $\sigma$ and any of the inputs. This is in general non-trivial, given that the model takes 178 inputs. Indeed, the only evident correlation exists with the value of the magnetic $L$ shell. Figure \ref{fig:dist_den2d} shows the two-dimensional histogram of $L$ and $\sigma$. The number of counts are normalized column-wise, that is for every value of $L$ the maximum is set equal to 1. The black dashed line follows the maximum number of counts as function of $L$. The uncertainty of the density estimation increases with increasing $L$, reaching a maximum for $L\sim 6$. This is consistent with the distribution of errors, when ordered as function of $L$ (Figure 3 in \citet{chu17}), reproduced here in Figure \ref{fig:err_den2d}, with the same format as before. Even though the mean value remains centered around zero, the spread of the errors increases with increasing $L$, hence resulting in larger uncertainties.
We conclude this Section by reproducing the result shown in Figure 6 of \citet{chu17}, where the authors have applied the DEN2D model to the moderate storm of 4 February 2011. Figure \ref{fig:storm_den2d} reproduces the estimated electron density at 6 different times, ranging from the quite time before the storm to the recovery phase after the storm. The color bar indicates the (log) electron density. Superimposed to each image we show the isolines of the standard deviation calculated with our new method. It is interesting to notice how $\sigma$ is as dynamic as the electron density. Being derived from the DEN2D model, the uncertainty is itself dependent on the time history of geomagnetic indexes and on geographical location.

\section{Estimation of chorus wave amplitude}\label{section:chorus}
\begin{figure}[ht]
\begin{center}
\centerline{\includegraphics[width=\columnwidth]{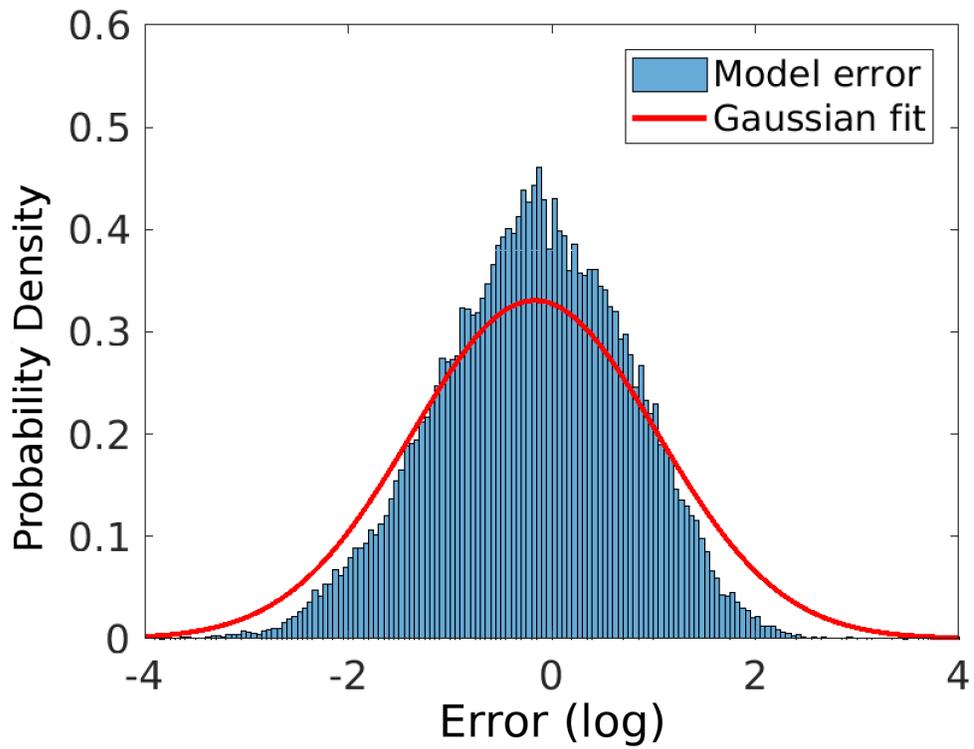}}
\caption{Density histogram of the errors of the model by \citet{agapitov18} (in logarithmic scale). The Gaussian fit is shown in red.}
\label{fig:error_agapitov}
\end{center}
\end{figure}

\begin{figure}[ht]
\begin{center}
\centerline{\includegraphics[width=\columnwidth]{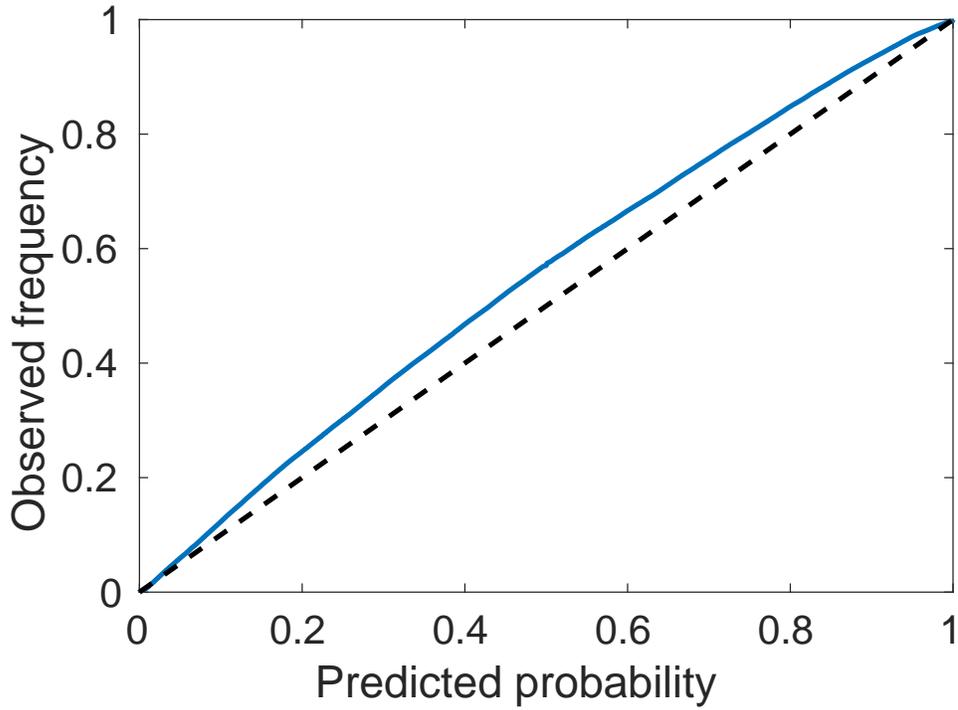}}
\caption{Reliability diagram for the probabilistic estimate of the chorus wave amplitude, based on the \citet{agapitov18} model. The black dashed line indicates perfect reliability.}
\label{fig:reliability_agapitov}
\end{center}
\end{figure}

\begin{figure}[ht]
\begin{center}
\centerline{\includegraphics[width=\columnwidth]{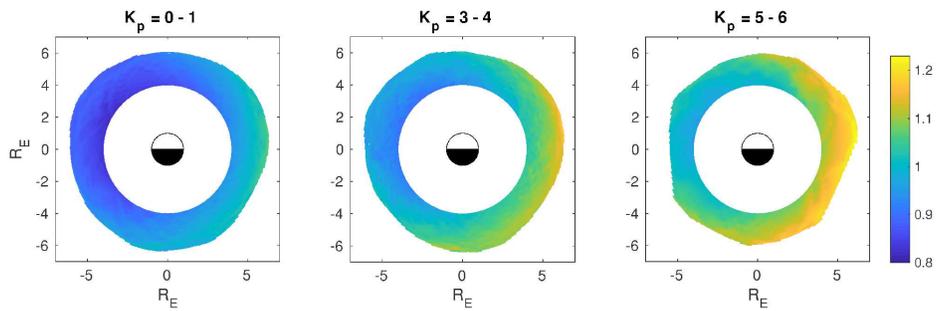}}
\caption{The standard deviation $\sigma$ estimated for the \citet{agapitov18} model (chorus wave amplitude), for three different ranges of the geomagnetic index Kp, as a function of different magnetic local time MLT and L shells.}
\label{fig:agapitov_std_kp}
\end{center}
\end{figure}

\begin{figure}[ht]
\begin{center}
\centerline{\includegraphics[width=\columnwidth]{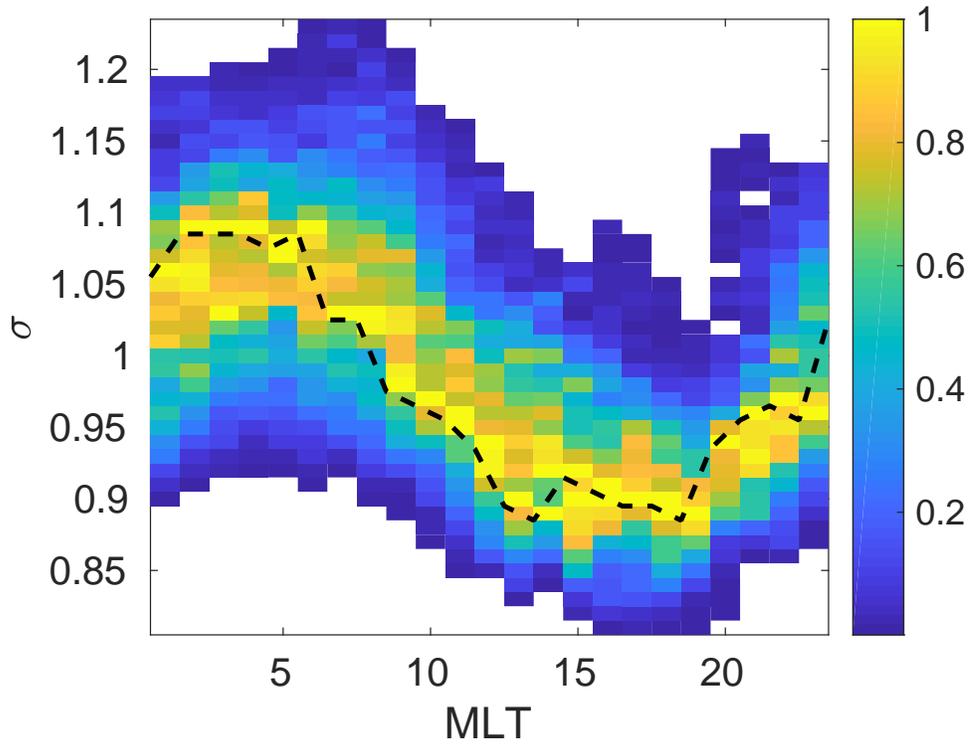}}
\caption{\citet{agapitov18} model (chorus wave amplitude). Two-dimensional histogram of standard deviation $\sigma$ as function of magnetic local time MLT. The number of counts are normalized column-wise: the maximum for each value of MLT is equal to 1.}
\label{fig:2D_hist_agapitov}
\end{center}
\end{figure}

Whistler-mode chorus waves play a crucial role for wave-particles interaction and particles scattering in the inner magnetosphere \citep{thorne10,camporeale15, camporeale15b}. The estimation of the wave amplitude is an important step in the calculation of pitch angle and energy diffusion by means of quasi-linear Fokker-Planck equations.
Recently, \citet{agapitov18} have presented an empirical model to estimate the chorus wave amplitude and wave normal angle distribution, derived from the statistical analysis of Cluster and Van Allen Probes VLF measurements. {The model takes as inputs the magnetic local time (MLT), the magnetic latitude $\lambda$, the value of the L-shell, and the geomagnetic index Kp (or Dst \citep{agapitov15}) providing the distribution of chorus wave amplitude and wave normal angle in the outer radiation belt (from the plasmapause to L=7) for all MLT values in the latitudinal range from -45 to 45 degrees. The model was developed in the polynomial form for chorus wave amplitude $B_w(\lambda,K_p) = a_{ij}\lambda^iK_p^i$ $(i,j=0,3)$ with the coefficients calculated based on Cluster STAFF-SA measurements in 2001-2011 \citep{agapitov15}, and with the coefficients updated making use of the combined Cluster observations and the recent Van Allen Probes VLF measurements \citep{agapitov18}.}
In order to apply our new method to the model of \citet{agapitov18}, we have produced an estimation of the chorus wave amplitude for the period 01-01-2015 to 12-30-2016, at one-minute resolution at the corresponding location of the Van Allen Probes spacecraft and the corresponding level of the geomagnetic activity. The ground-truth value is taken directly from the Van Allen Probes EMFISIS observations. Note that this time interval was not included in the original training of the model. This produced a total of 213,937 data points for which the model error was calculated. Since the wave amplitude can range within two orders of magnitude, the errors are in log scale. \\
Figure \ref{fig:error_agapitov} shows the histogram of the model error (computed as the difference between the logarithm of predictions and observations), compared with its Gaussian fit. Similarly to the model discussed in the previous Section, this model does not yield errors that are exactly log-normal distributed. This, however, does not affect the goodness of our uncertainty estimate, in terms of accuracy and reliability. 
As previously, we train our algorithm to estimate the standard deviation using the same inputs as the original model. The reliability diagram, calculated over the entire dataset, is shown in Figure \ref{fig:reliability_agapitov}. The largest mismatch, for a predicted probability equal to 50\%, is about 7\%, hence demonstrating that the model is very well calibrated.
Figure \ref{fig:agapitov_std_kp} shows the heat map of the standard deviation $\sigma$ at different locations $4<L<6.5$, and for different ranges of the geomagnetic index Kp (left panel: $Kp=[0-1]$; center panel: $Kp=[3-4]$; right panel: $Kp=[5-6]$). Not surprisingly, the largest uncertainties occur during storm-time, and in the pre-noon sector. 
Finally, Figure \ref{fig:2D_hist_agapitov} shows the two-dimensional histogram of the standard deviation $\sigma$, as function of the magnetic local time MLT. A column-wise normalization is applied, such that the maximum value along a constant MLT is equal to one. Consistently with the previous Figure, the largest uncertainties occur for MLT in the range 5-10. 

\section{Conclusions}\label{sec:conclusions}
The estimation of uncertainties associated with the output of deterministic models is a key element of any forecasting method. The standard approach for evaluating such uncertainties is to rely on time-consuming ensemble simulations. In this paper, we have introduced a novel methodology to estimate uncertainties that does not require running costly ensembles. The guiding principle behind our method is that the uncertainty of the output distribution, here represented by the standard deviation of a Gaussian centered around the values predicted by the deterministic model, should produce a probabilistic forecast that is both accurate and reliable (well calibrated). We have introduced a cost function that encodes the trade-off between accuracy and reliability for Gaussian distributions. The minimization of such Accuracy-Reliability cost function yields the optimal standard deviation $\sigma$.
The proposed method is ignorant with respect to the deterministic model it is applied to. In fact, it only requires the algebraic errors between predictions and true values, in order to be trained. A deep neural network is used to generate the unknown standard deviation for inputs other than the ones used for training.\\
We have shown experiments with synthetic data sets (for one and five-dimensional examples), that demonstrate how our method is able to learn the underlying functional dependence of the standard deviation, which is, in real-world problems, unknown. {These experiments also show how the method deals with cases when the underlying deterministic model contains a systematic error. In this cases, the reliability diagram represents a sanity check, indicating the presence of systematic errors. Indeed, it is understood that any problem with the underlying deterministic model is ultimately reflected in the reliability diagram.}\\
Finally we have applied the method to two recently developed models, relevant to space weather: the estimation of the electron density in the plasmasphere (Section \ref{section:dend2d}), and of the chorus wave amplitude (Section \ref{section:chorus}). In both cases, we use as inputs the same inputs employed in the original model. The probabilistic forecast produced with our method show excellent reliability diagrams, pinpointing the lack of a systematic bias in the original models.\\
Our code is available on the website \texttt{www.mlspaceweather.org} {and zenodo.org (doi:10.5281/zenodo.1485608)
} and we encourage the space weather community to produce probabilistic forecasts based on deterministic models, using our method.
{Finally, we point out that an interesting future extension to this method would be the case of multivariate outputs (in contrast to scalars). In that case, the definitions of CRPS and RS will need to account for co-variances between variables. }

\clearpage
\newpage

\clearpage
\acknowledgments
E.C. is partially funded by NWO-Vidi grant 639.072.716.\\
All data used to generate the results are available on the website \texttt{www.mlspaceweather.org} {and zenodo.org (doi:10.5281/zenodo.1485608)}

 \newpage

\listofchanges
%%%


\begin{thebibliography}{62}
\providecommand{\natexlab}[1]{#1}
\expandafter\ifx\csname urlstyle\endcsname\relax
  \providecommand{\doi}[1]{doi:\discretionary{}{}{}#1}\else
  \providecommand{\doi}{doi:\discretionary{}{}{}\begingroup
  \urlstyle{rm}\Url}\fi

\bibitem[{\textit{Agapitov et~al.}(2015)\textit{Agapitov, Artemyev, Mourenas,
  Mozer, and Krasnoselskikh}}]{agapitov15}
Agapitov, O., A.~Artemyev, D.~Mourenas, F.~Mozer, and V.~Krasnoselskikh (2015),
  Empirical model of lower band chorus wave distribution in the outer radiation
  belt, \textit{Journal of Geophysical Research: Space Physics},
  \textit{120}(12), 10--425.

\bibitem[{\textit{Agapitov et~al.}(2018)\textit{Agapitov, Mourenas, Artemyev,
  Mozer, Hospodarsky, Bonnell, and Krasnoselskikh}}]{agapitov18}
Agapitov, O., D.~Mourenas, A.~Artemyev, F.~Mozer, G.~Hospodarsky, J.~Bonnell,
  and V.~Krasnoselskikh (2018), Synthetic empirical chorus wave model from
  combined van allen probes and cluster statistics, \textit{Journal of
  Geophysical Research: Space Physics}, \textit{123}(1), 297--314.

\bibitem[{\textit{Aminalragia-Giamini et~al.}(2018)\textit{Aminalragia-Giamini,
  Sandberg, Papadimitriou, Daglis, and Jiggens}}]{aminalragia18}
Aminalragia-Giamini, S., I.~Sandberg, C.~Papadimitriou, I.~A. Daglis, and
  P.~Jiggens (2018), The virtual enhancements- solar proton event radiation
  (vesper) model, \textit{Journal of Space Weather and Space Climate},
  \textit{8}, A06.

\bibitem[{\textit{Anderson}(1996)}]{anderson96}
Anderson, J.~L. (1996), A method for producing and evaluating probabilistic
  forecasts from ensemble model integrations, \textit{Journal of Climate},
  \textit{9}(7), 1518--1530.

\bibitem[{\textit{Babu{\v{s}}ka et~al.}(2007)\textit{Babu{\v{s}}ka, Nobile, and
  Tempone}}]{babuvska07}
Babu{\v{s}}ka, I., F.~Nobile, and R.~Tempone (2007), A stochastic collocation
  method for elliptic partial differential equations with random input data,
  \textit{SIAM Journal on Numerical Analysis}, \textit{45}(3), 1005--1034.

\bibitem[{\textit{Barnes et~al.}(2007)\textit{Barnes, Leka, Schumer, and
  Della‐Rose}}]{barnes03}
Barnes, G., K.~D. Leka, E.~A. Schumer, and D.~J. Della‐Rose (2007),
  Probabilistic forecasting of solar flares from vector magnetogram data,
  \textit{Space Weather}, \textit{5}(9), \doi{10.1029/2007SW000317}.

\bibitem[{\textit{Barnes et~al.}(2016)\textit{Barnes, Leka, Schrijver, Colak,
  Qahwaji, Ashamari, Yuan, Zhang, McAteer, Bloomfield et~al.}}]{barnes16}
Barnes, G., K.~Leka, C.~Schrijver, T.~Colak, R.~Qahwaji, O.~Ashamari, Y.~Yuan,
  J.~Zhang, R.~McAteer, D.~Bloomfield, et~al. (2016), A comparison of flare
  forecasting methods. i. results from the “all-clear” workshop,
  \textit{The Astrophysical Journal}, \textit{829}(2), 89.

\bibitem[{\textit{Bishop}(1995)}]{bishop95}
Bishop, C.~M. (1995), \textit{Neural networks for pattern recognition}, Oxford
  university press.

\bibitem[{\textit{Bloomfield et~al.}(2012)\textit{Bloomfield, Higgins, McAteer,
  and Gallagher}}]{bloomfield12}
Bloomfield, D.~S., P.~A. Higgins, R.~J. McAteer, and P.~T. Gallagher (2012),
  Toward reliable benchmarking of solar flare forecasting methods, \textit{The
  Astrophysical Journal Letters}, \textit{747}(2), L41.

\bibitem[{\textit{Branke et~al.}(2008)\textit{Branke, Deb, and
  Miettinen}}]{branke08}
Branke, J., K.~Deb, and K.~Miettinen (2008), \textit{Multiobjective
  optimization: Interactive and evolutionary approaches}, vol. 5252, Springer
  Science \& Business Media.

\bibitem[{\textit{Brier}(1950)}]{brier50}
Brier, G.~W. (1950), Verification of forecasts expressed in terms of
  probability, \textit{Monthly Weather Review}, \textit{78}(1), 1--3.

\bibitem[{\textit{Br{\"o}cker and Smith}(2007)}]{brocker07}
Br{\"o}cker, J., and L.~A. Smith (2007), Scoring probabilistic forecasts: The
  importance of being proper, \textit{Weather and Forecasting}, \textit{22}(2),
  382--388.

\bibitem[{\textit{Bussy‐Virat and Ridley}(2014)}]{Bussy‐Virat14}
Bussy‐Virat, C.~D., and A.~J. Ridley (2014), Predictions of the solar wind
  speed by the probability distribution function model, \textit{Space Weather},
  \textit{12}(6), 337--353, \doi{10.1002/2014SW001051}.

\bibitem[{\textit{Caflisch}(1998)}]{caflisch98}
Caflisch, R.~E. (1998), Monte carlo and quasi-monte carlo methods, \textit{Acta
  numerica}, \textit{7}, 1--49.

\bibitem[{\textit{Camporeale}(2015)}]{camporeale15}
Camporeale, E. (2015), Resonant and nonresonant whistlers-particle interaction
  in the radiation belts, \textit{Geophysical Research Letters},
  \textit{42}(9), 3114--3121.

\bibitem[{\textit{Camporeale and Zimbardo}(2015)}]{camporeale15b}
Camporeale, E., and G.~Zimbardo (2015), Wave-particle interactions with
  parallel whistler waves: Nonlinear and time-dependent effects revealed by
  particle-in-cell simulations, \textit{Physics of Plasmas}, \textit{22}(9),
  092,104.

\bibitem[{\textit{Camporeale et~al.}(2016)\textit{Camporeale, Shprits,
  Chandorkar, Drozdov, and Wing}}]{camporeale16}
Camporeale, E., Y.~Shprits, M.~Chandorkar, A.~Drozdov, and S.~Wing (2016), On
  the propagation of uncertainties in radiation belt simulations, \textit{Space
  Weather}, \textit{14}(11), 982--992.

\bibitem[{\textit{Camporeale et~al.}(2017)\textit{Camporeale, Agnihotri, and
  Rutjes}}]{camporeale17}
Camporeale, E., A.~Agnihotri, and C.~Rutjes (2017), Adaptive selection of
  sampling points for uncertainty quantification, \textit{International Journal
  for Uncertainty Quantification}, \textit{7}(4).

\bibitem[{\textit{Camporeale et~al.}(2018{\natexlab{a}})\textit{Camporeale,
  Wing, Johnson, Jackman, and McGranaghan}}]{camporeale18}
Camporeale, E., S.~Wing, J.~Johnson, C.~Jackman, and R.~McGranaghan
  (2018{\natexlab{a}}), Space weather in the machine learning era: a
  multi-disciplinary approach, \textit{Space Weather}.

\bibitem[{\textit{Camporeale et~al.}(2018{\natexlab{b}})\textit{Camporeale,
  Wing, and Johnson}}]{camporeale18b}
Camporeale, E., S.~Wing, and J.~R. Johnson (2018{\natexlab{b}}),
  \textit{Machine Learning Techniques for Space Weather}, Elsevier.

\bibitem[{\textit{Car{\`e} and Camporeale}(2018)}]{care18}
Car{\`e}, A., and E.~Camporeale (2018), Regression, in \textit{Machine Learning
  Techniques for Space Weather}, pp. 71--112, Elsevier.

\bibitem[{\textit{Chandorkar et~al.}(2017)\textit{Chandorkar, Camporeale, and
  Wing}}]{chandorkar17}
Chandorkar, M., E.~Camporeale, and S.~Wing (2017), Probabilistic forecasting of
  the disturbance storm time index: An autoregressive gaussian process
  approach, \textit{Space Weather}, \textit{15}(8), 1004--1019,
  \doi{10.1002/2017SW001627}.

\bibitem[{\textit{Chu et~al.}(2017)\textit{Chu, Bortnik, Li, Ma, Angelopoulos,
  and Thorne}}]{chu17}
Chu, X., J.~Bortnik, W.~Li, Q.~Ma, V.~Angelopoulos, and R.~Thorne (2017),
  Erosion and refilling of the plasmasphere during a geomagnetic storm modeled
  by a neural network, \textit{Journal of Geophysical Research: Space Physics}.

\bibitem[{\textit{Gallagher et~al.}(2002)\textit{Gallagher, Moon, and
  Wang}}]{gallagher02}
Gallagher, P.~T., Y.-J. Moon, and H.~Wang (2002), Active-region monitoring and
  flare forecasting--i. data processing and first results, \textit{Solar
  Physics}, \textit{209}(1), 171--183.

\bibitem[{\textit{Genz}(1984)}]{genz84}
Genz, A. (1984), Testing multidimensional integration routines, in
  \textit{Proc. Of International Conference on Tools, Methods and Languages for
  Scientific and Engineering Computation}, pp. 81--94, Elsevier North-Holland,
  Inc., New York, NY, USA.

\bibitem[{\textit{Ghahramani}(2015)}]{ghahramani15}
Ghahramani, Z. (2015), Probabilistic machine learning and artificial
  intelligence, \textit{Nature}, \textit{521}(7553), 452.

\bibitem[{\textit{Gneiting et~al.}(2005)\textit{Gneiting, Raftery,
  Westveld~III, and Goldman}}]{gneiting05}
Gneiting, T., A.~E. Raftery, A.~H. Westveld~III, and T.~Goldman (2005),
  Calibrated probabilistic forecasting using ensemble model output statistics
  and minimum crps estimation, \textit{Monthly Weather Review},
  \textit{133}(5), 1098--1118.

\bibitem[{\textit{Gneiting et~al.}(2007)\textit{Gneiting, Balabdaoui, and
  Raftery}}]{gneiting07}
Gneiting, T., F.~Balabdaoui, and A.~E. Raftery (2007), Probabilistic forecasts,
  calibration and sharpness, \textit{Journal of the Royal Statistical Society:
  Series B (Statistical Methodology)}, \textit{69}(2), 243--268.

\bibitem[{\textit{Goldberg et~al.}(1998)\textit{Goldberg, Williams, and
  Bishop}}]{goldberg98}
Goldberg, P.~W., C.~K. Williams, and C.~M. Bishop (1998), Regression with
  input-dependent noise: A gaussian process treatment, in \textit{Advances in
  neural information processing systems}, pp. 493--499.

\bibitem[{\textit{Gruet et~al.}(2018)\textit{Gruet, Chandorkar, Sicard, and
  Camporeale}}]{gruet18}
Gruet, M., M.~Chandorkar, A.~Sicard, and E.~Camporeale (2018), Multiple hours
  ahead forecast of the dst index using a combination of long short-term memory
  neural network and gaussian process, \textit{Space Weather}, \textit{under
  review}.

\bibitem[{\textit{Hamill}(1997)}]{hamill97}
Hamill, T.~M. (1997), Reliability diagrams for multicategory probabilistic
  forecasts, \textit{Weather and forecasting}, \textit{12}(4), 736--741.

\bibitem[{\textit{Hamill}(2001)}]{hamill01}
Hamill, T.~M. (2001), Interpretation of rank histograms for verifying ensemble
  forecasts, \textit{Monthly Weather Review}, \textit{129}(3), 550--560.

\bibitem[{\textit{Hersbach}(2000)}]{hersbach00}
Hersbach, H. (2000), Decomposition of the continuous ranked probability score
  for ensemble prediction systems, \textit{Weather and Forecasting},
  \textit{15}(5), 559--570.

\bibitem[{\textit{Kahler and Ling}(2015)}]{kahler15}
Kahler, S.~W., and A.~Ling (2015), Dynamic sep event probability forecasts,
  \textit{Space Weather}, \textit{13}(10), 665--675,
  \doi{10.1002/2015SW001222}.

\bibitem[{\textit{Kennedy and O'Hagan}(2001)}]{kennedy01}
Kennedy, M.~C., and A.~O'Hagan (2001), Bayesian calibration of computer models,
  \textit{Journal of the Royal Statistical Society: Series B (Statistical
  Methodology)}, \textit{63}(3), 425--464.

\bibitem[{\textit{Kersting et~al.}(2007)\textit{Kersting, Plagemann, Pfaff, and
  Burgard}}]{kersting07}
Kersting, K., C.~Plagemann, P.~Pfaff, and W.~Burgard (2007), Most likely
  heteroscedastic gaussian process regression, in \textit{Proceedings of the
  24th international conference on Machine learning}, pp. 393--400, ACM.

\bibitem[{\textit{Lee et~al.}(2012)\textit{Lee, Moon, Lee, Lee, and
  Na}}]{lee12}
Lee, K., Y.-J. Moon, J.-Y. Lee, K.-S. Lee, and H.~Na (2012), Solar flare
  occurrence rate and probability in terms of the sunspot classification
  supplemented with sunspot area and its changes, \textit{Solar Physics},
  \textit{281}(2), 639--650.

\bibitem[{\textit{Luhmann et~al.}(2017)\textit{Luhmann, Mays, Odstrcil, Li,
  Bain, Lee, Galvin, Mewaldt, Cohen, Leske, Larson, and Futaana}}]{Luhmann17}
Luhmann, J.~G., M.~L. Mays, D.~Odstrcil, Y.~Li, H.~Bain, C.~O. Lee, A.~B.
  Galvin, R.~A. Mewaldt, C.~M.~S. Cohen, R.~A. Leske, D.~Larson, and Y.~Futaana
  (2017), Modeling solar energetic particle events using enlil heliosphere
  simulations, \textit{Space Weather}, \textit{15}(7), 934--954,
  \doi{10.1002/2017SW001617}.

\bibitem[{\textit{Matheson and Winkler}(1976)}]{matheson76}
Matheson, J.~E., and R.~L. Winkler (1976), Scoring rules for continuous
  probability distributions, \textit{Management science}, \textit{22}(10),
  1087--1096.

\bibitem[{\textit{McPherron and Siscoe}(2004)}]{mcPherron04}
McPherron, R.~L., and G.~Siscoe (2004), Probabilistic forecasting of
  geomagnetic indices using solar wind air mass analysis, \textit{Space
  Weather}, \textit{2}(1), \doi{10.1029/2003SW000003}.

\bibitem[{\textit{Miyoshi and Kataoka}(2008)}]{miyoshi08}
Miyoshi, Y., and R.~Kataoka (2008), Probabilistic space weather forecast of the
  relativistic electron flux enhancement at geosynchronous orbit,
  \textit{Journal of Atmospheric and Solar-Terrestrial Physics},
  \textit{70}(2-4), 475--481.

\bibitem[{\textit{M\"{o}stl et~al.}(2017)\textit{M\"{o}stl, Amerstorfer,
  Palmerio, Isavnin, Farrugia, Lowder, Winslow, Donnerer, Kilpua, and
  Boakes}}]{mostl17}
M\"{o}stl, C., T.~Amerstorfer, E.~Palmerio, A.~Isavnin, C.~J. Farrugia,
  C.~Lowder, R.~M. Winslow, J.~M. Donnerer, E.~K.~J. Kilpua, and P.~D. Boakes
  (2017), Forward modeling of coronal mass ejection flux ropes in the inner
  heliosphere with 3dcore, \textit{Space Weather}, \textit{16}(3), 216--229.

\bibitem[{\textit{Murphy and Winkler}(1992)}]{murphy92}
Murphy, A.~H., and R.~L. Winkler (1992), Diagnostic verification of probability
  forecasts, \textit{International Journal of Forecasting}, \textit{7}(4),
  435--455.

\bibitem[{\textit{Murphy}(2012)}]{murphy12}
Murphy, K.~P. (2012), Machine learning: A probabilistic perspective. adaptive
  computation and machine learning.

\bibitem[{\textit{Murray et~al.}(2017)\textit{Murray, Bingham, Sharpe, and
  Jackson}}]{Murray17}
Murray, S.~A., S.~Bingham, M.~Sharpe, and D.~R. Jackson (2017), Flare
  forecasting at the met office space weather operations centre, \textit{Space
  Weather}, \textit{15}(4), 577--588, \doi{10.1002/2016SW001579}.

\bibitem[{\textit{Napoletano et~al.}(2018)\textit{Napoletano, Forte, Del~Moro,
  Pietropaolo, Giovannelli, and Berrilli}}]{napoletano18}
Napoletano, G., R.~Forte, D.~Del~Moro, E.~Pietropaolo, L.~Giovannelli, and
  F.~Berrilli (2018), A probabilistic approach to the drag-based model,
  \textit{arXiv preprint arXiv:1801.04201}.

\bibitem[{\textit{Owens and Riley}(2017)}]{owens17}
Owens, M.~J., and P.~Riley (2017), Probabilistic solar wind forecasting using
  large ensembles of near‐sun conditions with a simple one‐dimensional
  “upwind” scheme, \textit{Space Weather}, \textit{15}(11), 1461--1474,
  \doi{10.1002/2017SW001679}.

\bibitem[{\textit{Papaioannou et~al.}(2015)\textit{Papaioannou, Anastasiadis,
  Sandberg, Georgoulis, Tsiropoula, Tziotziou, Jiggens, and
  Hilgers}}]{papaioannou15}
Papaioannou, A., A.~Anastasiadis, I.~Sandberg, M.~Georgoulis, G.~Tsiropoula,
  K.~Tziotziou, P.~Jiggens, and A.~Hilgers (2015), A novel forecasting system
  for solar particle events and flares (forspef), in \textit{Journal of
  Physics: Conference Series}, vol. 632, p. 012075, IOP Publishing.

\bibitem[{\textit{Pinson et~al.}(2010)\textit{Pinson, McSharry, and
  Madsen}}]{pinson10}
Pinson, P., P.~McSharry, and H.~Madsen (2010), Reliability diagrams for
  non-parametric density forecasts of continuous variables: Accounting for
  serial correlation, \textit{Quarterly Journal of the Royal Meteorological
  Society}, \textit{136}(646), 77--90.

\bibitem[{\textit{Prikryl et~al.}(2012)\textit{Prikryl, Jayachandran, Mushini,
  and Richardson}}]{prikryl2012}
Prikryl, P., P.~Jayachandran, S.~Mushini, and I.~Richardson (2012), Toward the
  probabilistic forecasting of high-latitude gps phase scintillation,
  \textit{Space Weather}, \textit{10}(8).

\bibitem[{\textit{Riley and Love}(2016)}]{riley16}
Riley, P., and J.~J. Love (2016), Extreme geomagnetic storms: Probabilistic
  forecasts and their uncertainties, \textit{Space Weather}, \textit{15}(1),
  53--64, \doi{10.1002/2016SW001470}.

\bibitem[{\textit{Sharpe and Murray}(2017)}]{Sharpe17}
Sharpe, M.~A., and S.~A. Murray (2017), Verification of space weather forecasts
  issued by the met office space weather operations centre, \textit{Space
  Weather}, \textit{15}(10), 1383--1395, \doi{10.1002/2017SW001683}.

\bibitem[{\textit{Smith}(2013)}]{smith13}
Smith, R.~C. (2013), \textit{Uncertainty quantification: theory,
  implementation, and applications}, vol.~12, Siam.

\bibitem[{\textit{Thorne}(2010)}]{thorne10}
Thorne, R.~M. (2010), Radiation belt dynamics: The importance of wave-particle
  interactions, \textit{Geophysical Research Letters}, \textit{37}(22).

\bibitem[{\textit{Weigend and Nix}(1994)}]{weigend94}
Weigend, A.~S., and D.~A. Nix (1994), Predictions with confidence intervals
  (local error bars), in \textit{Proceedings of the international conference on
  neural information processing}, pp. 847--852.

\bibitem[{\textit{Wheatland}(2004)}]{wheatland04}
Wheatland, M. (2004), A bayesian approach to solar flare prediction,
  \textit{The Astrophysical Journal}, \textit{609}(2), 1134.

\bibitem[{\textit{Wilks}(2011)}]{wilks11}
Wilks, D.~S. (2011), \textit{Statistical methods in the atmospheric sciences},
  vol. 100, Academic press.

\bibitem[{\textit{Williams}(1996)}]{williams96}
Williams, P.~M. (1996), Using neural networks to model conditional multivariate
  densities, \textit{Neural Computation}, \textit{8}(4), 843--854.

\bibitem[{\textit{Xiu}(2010)}]{xiu10}
Xiu, D. (2010), \textit{Numerical methods for stochastic computations: a
  spectral method approach}, Princeton university press.

\bibitem[{\textit{Xiu and Karniadakis}(2002)}]{xiu02}
Xiu, D., and G.~E. Karniadakis (2002), The wiener--askey polynomial chaos for
  stochastic differential equations, \textit{SIAM journal on scientific
  computing}, \textit{24}(2), 619--644.

\bibitem[{\textit{Yuan and Wahba}(2004)}]{yuan04}
Yuan, M., and G.~Wahba (2004), Doubly penalized likelihood estimator in
  heteroscedastic regression, \textit{Statistics \& probability letters},
  \textit{69}(1), 11--20.

\bibitem[{\textit{Zhang and Moldwin}(2014)}]{zhang14}
Zhang, X., and M.~B. Moldwin (2014), Probabilistic forecasting analysis of
  geomagnetic indices for southward imf events, \textit{Space Weather},
  \textit{13}(3), 130--140, \doi{10.1002/2014SW001113}.

\end{thebibliography}
\end{document}